\documentclass[twocolumn]{aastex631}
\accepted{to the PASP, August 29, 2024}

\begin{document}

\title{Design and Performance of the Upgraded Mid-InfraRed Spectrometer and Imager (MIRSI) on the NASA Infrared Telescope Facility}

\author[0000-0002-5599-4650]{Joseph L. Hora}
\affiliation{Center for Astrophysics $|$ Harvard \& Smithsonian,
60 Garden Street,
Cambridge, MA 02138, USA}

\author[0000-0003-4580-3790]{David E. Trilling}
\affiliation{Department of Astronomy and Planetary Science, Northern Arizona University, Flagstaff, AZ 86011, USA}

\author[0000-0002-2601-6954]{Andy J. L\'{o}pez-Oquendo}
\affiliation{Department of Astronomy and Planetary Science, Northern Arizona University, Flagstaff, AZ 86011, USA}

\author{Howard A. Smith}
\affiliation{Center for Astrophysics $|$ Harvard \& Smithsonian,
60 Garden Street,
Cambridge, MA 02138, USA}

\author[0000-0002-8132-778X]{Michael Mommert}
\affiliation{Stuttgart University of Applied Sciences, Stuttgart, Germany}
\affiliation{Department of Astronomy and Planetary Science, Northern Arizona University, Flagstaff, AZ 86011, USA}
\affiliation{Lowell Observatory, 1400 West Mars Hill Road, Flagstaff, AZ 86001, USA}

\author[0000-0001-6765-6336]{Nicholas Moskovitz}
\affiliation{Lowell Observatory, 1400 West Mars Hill Road, Flagstaff, AZ 86001, USA}

\author{Chris Foster}
\affiliation{Infrared Laboratories, Inc., 1808 E. 17th St, Tucson, Arizona, 85719, USA }

\author[0000-0002-8293-1428]{Michael S. Connelley}
\affiliation{University of Hawai`i, 640 A'ohōkū Place, Hilo, HI 96720, USA}

\author{Charles Lockhart}
\affiliation{University of Hawai`i, 640 A'ohōkū Place, Hilo, HI 96720, USA}
\author[0000-0002-3165-159X]{John T. Rayner}
\affiliation{University of Hawai`i, 2680 Woodlawn Dr., Honolulu, HI 96822, USA}

\author[0000-0003-4191-6536]{Schelte J. Bus}
\affiliation{University of Hawai`i, 640 A'ohōkū Place, Hilo, HI 96720, USA}
\author{Darryl Watanabe}
\affiliation{University of Hawai`i, 640 A'ohōkū Place, Hilo, HI 96720, USA}
\author{Lars Bergknut}
\affiliation{University of Hawai`i, 640 A'ohōkū Place, Hilo, HI 96720, USA}
\author{Morgan Bonnet}
\affiliation{University of Hawai`i, 640 A'ohōkū Place, Hilo, HI 96720, USA}
\author[0000-0001-8136-9704]{Alan Tokunaga}
\affiliation{University of Hawai`i, 2680 Woodlawn Dr., Honolulu, HI 96822, USA}



\begin{abstract}

We describe the new design and current performance of the Mid-InfraRed Spectrometer and Imager (MIRSI) on the NASA Infrared Telescope Facility (IRTF). The system has been converted from a liquid nitrogen/liquid helium cryogen system to one that uses a closed-cycle cooler, which allows it to be kept on the telescope at operating temperature and available for observing on short notice, requiring less effort by the telescope operators and day crew to maintain operating temperature. Several other enhancements have been completed, including new detector readout electronics, an IRTF-style standard instrument user interface, new stepper motor driver electronics, and an optical camera that views the same field as the mid-IR instrument using a cold dichroic mirror, allowing for guiding and/or simultaneous optical imaging. The instrument performance is presented, both with an engineering-grade array used from 2021-2023, and a science-grade array installed in the fall of 2023. Some sample astronomical results are also shown. The upgraded MIRSI is a facility instrument at the IRTF available to all users.
\end{abstract}

\keywords{Astronomical instrumentation (799) --- Infrared astronomy (786) --- Spectrometers (1554)}


\section{Introduction}
The Mid-Infrared Spectrometer and Imager (MIRSI) was developed at Boston University by a team led by Lynne Deutsch \citep{2003SPIE.4841..106D, 2008PASP..120.1271K} and was used from 2002 -- 2011 on the NASA Infrared Telescope Facility (IRTF). MIRSI was used to make observations in the 2 – 25 \micron\ wavelength range of asteroids, planets, and comets \citep[e.g.,][]{2014MNRAS.439.3357Y,2023NatAs...7..190O,2011ApJ...734L...1M}, as well as observations for non-solar system science programs such as photodissociation regions \citep{2006ApJ...637..823K}, eclipsing binary stars \citep{2011AJ....142..174S}, and high mass protostars \citep{2009ApJ...699.1300W}. MIRSI was scheduled for all or part of 425 separate nights during this period, peaking at 46 nights in the 2005A semester. Over 65 publications were based in part on MIRSI observations\footnote{For a partial list of publications see \url{https://cfa.harvard.edu/mirsi/}}.

\subsection{Instrument Operation Issues}
The nominal orientation of the original MIRSI dewar on the telescope was with the window pointed upwards toward the incoming telescope beam and the cryogen cans on their side, extending horizontally away from the optical axis. This allowed for filling the cryogen without removing the instrument from its mount on the telescope, simplifying and speeding up the process. However, this orientation put more stress on the G10 supports that held the cans and radiation shields in place, and eventually these partially failed, allowing the outer shield to make contact with the inner LHe shield and reduced the dewar hold time. In 2010, the system was disassembled and the supports replaced with thicker structures which could properly support the weight of the shields.

Near the end of its original period of use, the instrument became more difficult to maintain due to its aging custom electronics and computer interface and control system, and more expensive to operate because of the high cost of supplying liquid helium (LHe) on Maunakea. The latter factor led to the instrument being used infrequently in discrete blocks of time to minimize the number of system cooldowns and LHe usage. These runs had to be planned well in advance to arrange for adequate LHe, and several times planned observing runs had to be cancelled due to delays in helium delivery by the supplier. During the last observing run, the detector array was damaged and rendered non-functional due to operating at higher than nominal temperatures. The LHe cryogen had boiled off faster than expected, and the original electronics did not have adequate safeguards to automatically turn off the array power if the operating temperature limits were exceeded. In addition, one of the readout electronics boards had failed, and replacement boards were not readily available.

\subsection{MIRSI Upgrade Science Goals}
In 2014 we proposed to the NASA Near-Earth Object Observations (NEOO) program to determine the diameters and albedos of 750 near-earth objects (NEOs) with observations from the IRTF with MIRSI. Most NEOs are discovered in ground-based optical surveys, and their diameters are very uncertain due to their unknown albedos which can vary from 0.02 to 0.42 or higher, depending on the asteroid class \citep{2011AJ....142...85T}.  Previous thermal IR observations with Spitzer/IRAC \citep{2010AJ....140..770T} and the NEOWISE mission \citep{2011ApJ...743..156M} had proven to be very effective in quickly characterizing NEOs. However, Spitzer had moved well away from the earth at that time, and NEOWISE's survey pattern and sensitivity often did not allow it to detect small newly-discovered NEOs. \label{MIRSIupgrade} Now that Spitzer and NEOWISE are no longer in operation, the IRTF is even more essential to fill this need.

It was clear at that point that MIRSI needed major work in order to become operational again, and that it could not operate in the same way as before with liquid cryogens. For the proposed NEO program, we would need to observe for a few hours every week or two, and wanted to focus on recently discovered NEOs that were still close enough to the earth to detect in the mid-IR. Some of these observations would require target of opportunity-style scheduling where we would need to observe high-priority objects within a matter of hours of their discovery. Clearly this was not economically or logistically feasible with liquid cryogens, so MIRSI would have to be converted to a closed-cycle cooler system. In addition, simultaneous optical measurements would be important in helping to constrain the albedo and diameter determination, and also allow for light curve measurements and guiding on the source which might not be visible in individual mid-IR images.

We describe in this paper the design and implementation of the  modifications performed to the MIRSI system to upgrade the cryogenic system to a closed-cycle cooler, and its performance as measured on the IRTF using an engineering-grade array from 2021-2023. We show some sample science images from the new system to demonstrate some of MIRSI's capabilities. In a companion paper \citep{Lopez-Oquendo2024}, we present results from the first observations of NEOs with the upgraded system.

\begin{figure}
    \centering
    \includegraphics[width=3in]{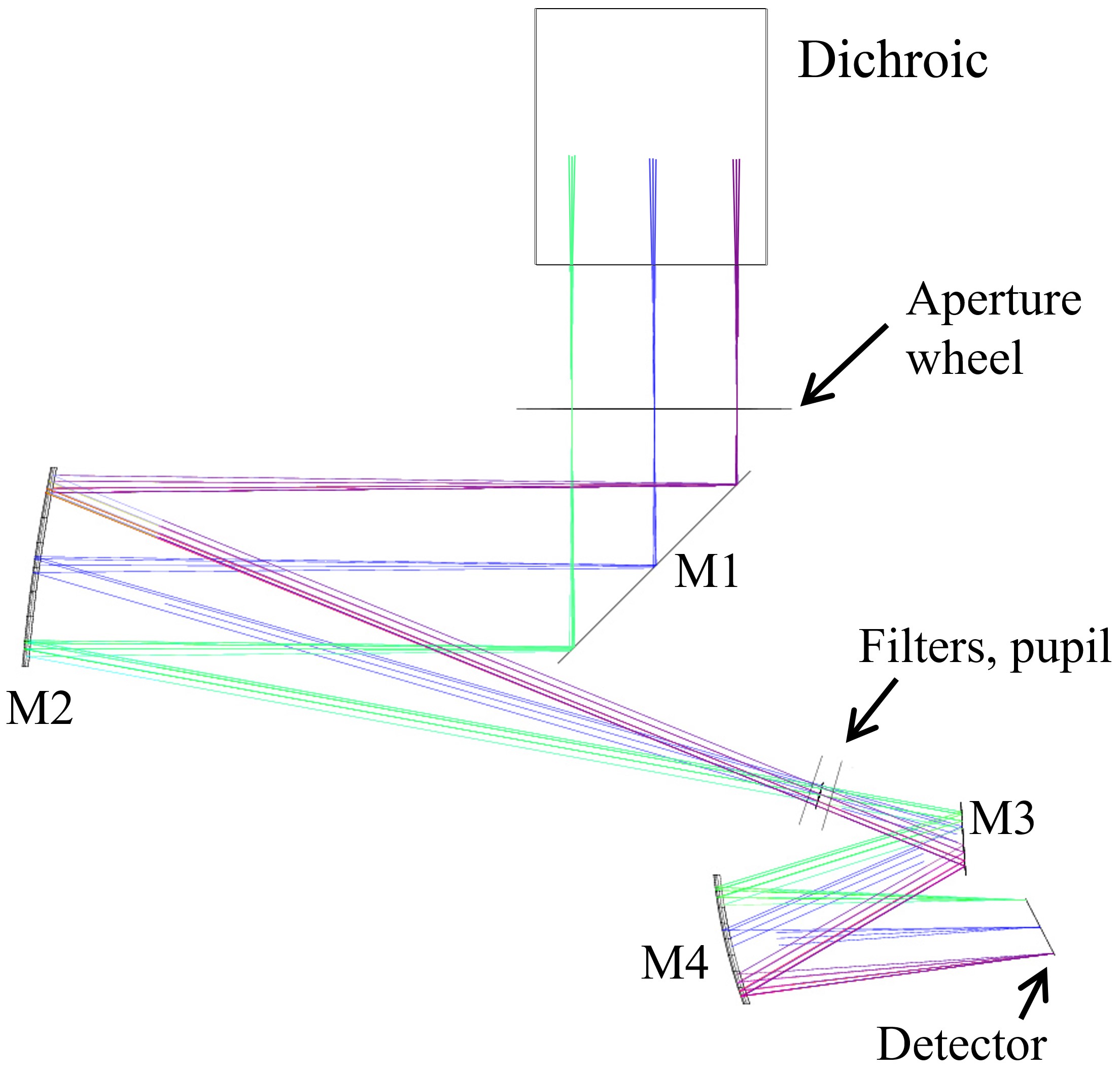}
    \vskip 0.1in
    \includegraphics[width=3.35in]{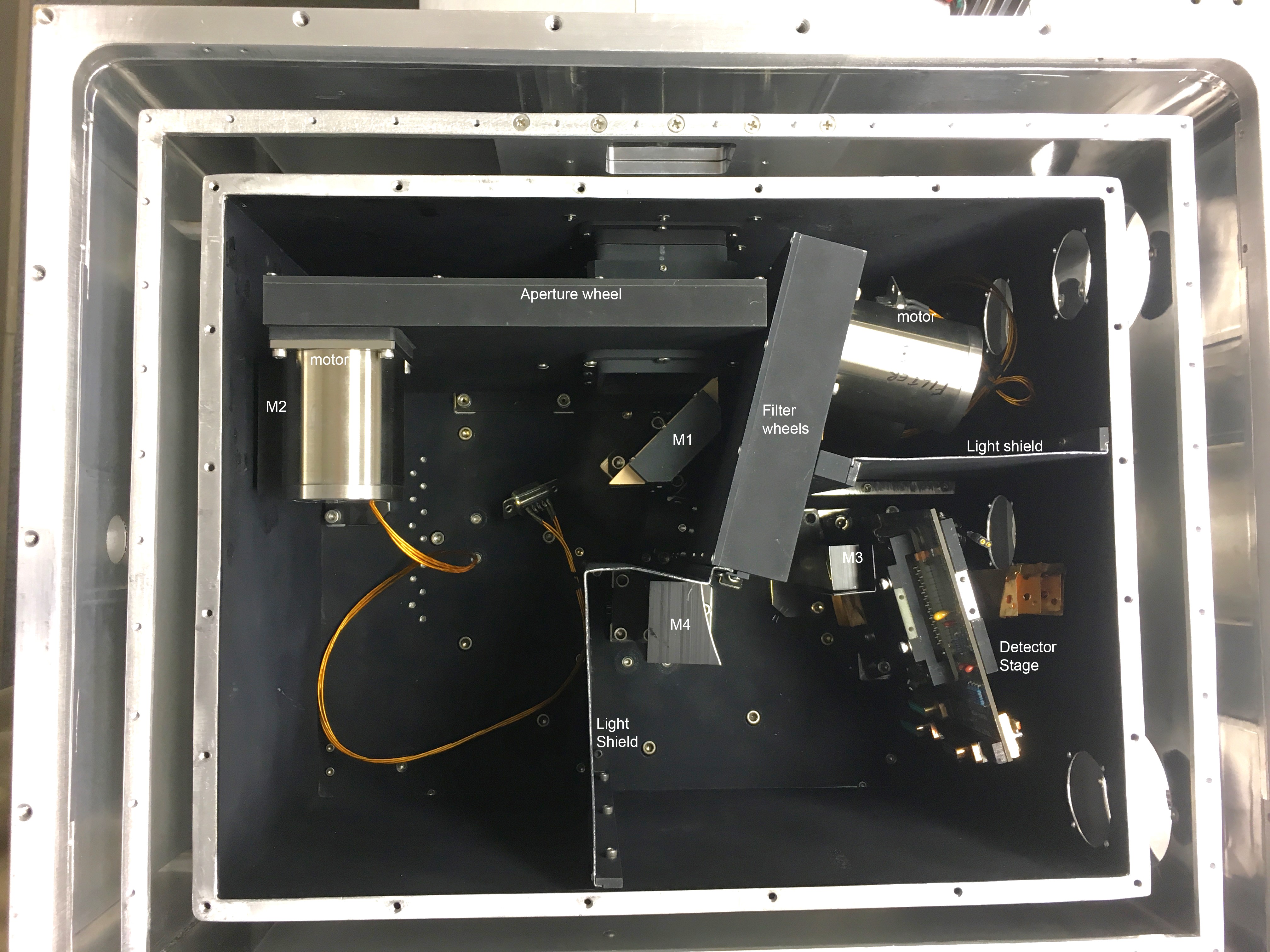}
    \caption{The top figure shows a ray-tracing of the optical path through the MIRSI dewar. The optical prescription of the mirrors is given in \citet{2008PASP..120.1271K}.  In the current instrument configuration, this view is looking up towards the telescope. The dichroic reflects the IR light coming perpendicular to the page into the optical system that is in the plane of the page. The lower image is a photo of the  inner part of the dewar in the same orientation, with some labels overlaid on the image. At the bottom one can see the optical plate to which is attached the inner light shield and the optical and mechanical components. The cover plates of the light shields have been removed to expose the mirrors and array mount in this photo. The electrical connectors are not installed, their through-holes are seen on the right side of the shields.
}
    \label{fig:optdesign}
\end{figure}
\begin{figure*}
    \centering
    \includegraphics[width=3in]{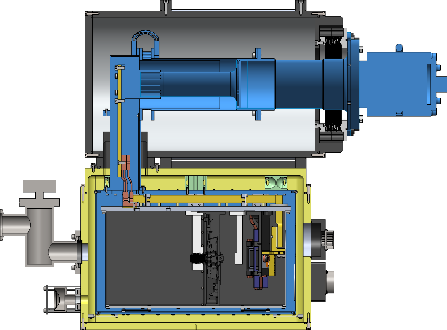}
    \includegraphics[width=3in]{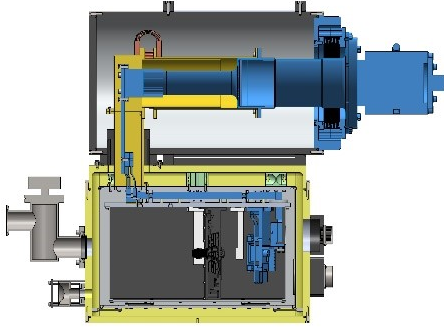}
    \caption{Two cutaway views of the dewar cooling design. The drawing on the left shows Stage 1 of the cold head (which cools the outer shield) in the blue color. The drawing on the right shows (in blue) the thermal path from Stage 2 of the cold head, which is entirely enclosed within the stage 1 path. The Stage 2 section includes the inner radiation shield and the camera optics \& filter wheels. The blue parts in the lower cryostat show the cooling path to the array stage and mount, which cools the detector. The optics mounting plate and inner shields are connected to Stage 2 through a separate thermal path, and reach temperatures $\sim$8~K.}
    \label{fig:cryostages}
\end{figure*}

\begin{figure}
    \centering
    \includegraphics[width = 0.45\textwidth]{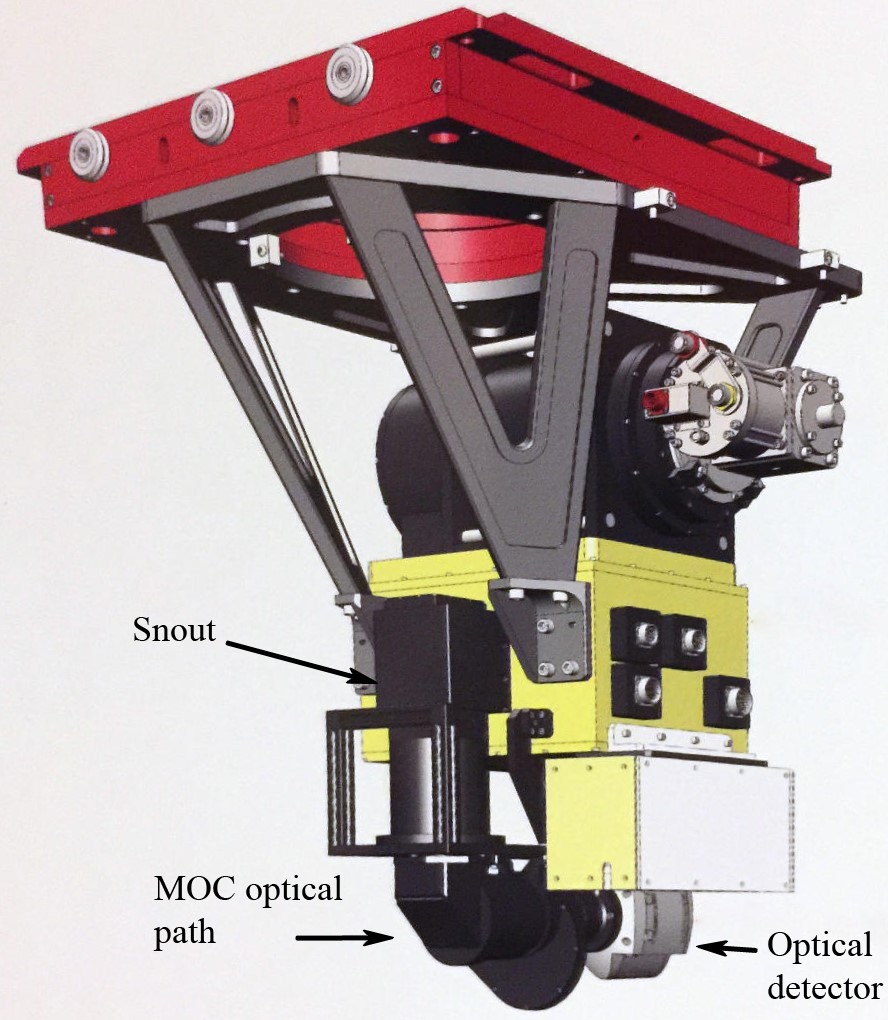}
    \caption{CAD rendering of MIRSI and the instrument mount on the IRTF. The red plate at the top is part of the IRTF trolley system to move the instrument into and away from the Cassegrain focus. Three V-shaped struts attach to the dewar at three points. The black section at the top is the cryocooler cold head mount which replaced the cryogen cans of the old system, and the gold section houses the original MIRSI optics and detector. Below this box is the detector readout electronics, and the optical camera (MIRSI Optical Camera or MOC; see \S \ref{sec:MOC}) with its filter wheel. The dewar snout and optical camera lightpath are in the black box and tube in the bottom front.}
    \label{fig:CADmount}
\end{figure}

\begin{figure}
    \centering
    \includegraphics[width = 3.3in]{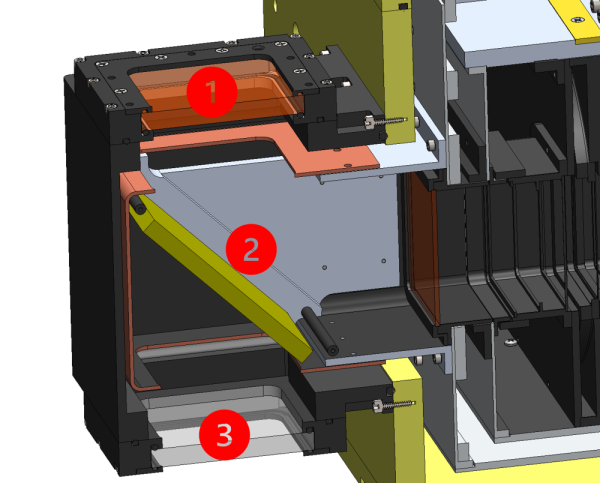}
    \caption{CAD cutaway rendering of the MIRSI ``snout" which holds the IR/optical dewar entrance window (1), the dichroic mirror (2), and the optical exit window (3), below which is mounted the optical camera. The dichroic holder and surrounding baffling is connected to the outer shield, the dark grey baffling on the right is connected to the inner shield. The aperture wheel is at the extreme right of the image, in the dark grey box.}
    \label{fig:snout}
\end{figure}

\section{MIRSI Upgrades}
\subsection{MIRSI description}
MIRSI was designed to observe at the IRTF at high spatial resolution with background-limited sensitivity. The system was optimized to acquire images and low-resolution grism spectra within the 8–14 and 17–26 \micron\ atmospheric windows. It also has K and M filters available in the 2 - 5~\micron\ range. A full description of the original MIRSI system can be found in \citet{2003SPIE.4841..106D} and \citet{2008PASP..120.1271K}. The optical layout and inner dewar with the optics components, mechanisms, and array mount are shown in Figure \ref{fig:optdesign}. This section is mostly unchanged in the new system, with the exception of a new array mount and connection to a cold finger from the second stage of the cooler, and replacing the filter position sensor mechanical switches with Hall effect sensors. The all-reflective design using gold-coated diamond-turned aluminum mirrors provides high throughput and excellent optical quality over the full wavelength range. 

A major change in the configuration of MIRSI is that a dichroic fold mirror has been added to reflect the IR light into the camera optics, allowing the light shortward of 1.5~\micron\ to pass through to an optical camera (described in \S\ref{sec:MOC}). The CAD renderings in Figures~\ref{fig:cryostages} and \ref{fig:CADmount} show the system in its nominal configuration with the telescope at zenith, with the optical plate horizontal and the optical components mounted below. The telescope beam enters through a ZnSe window which is mounted on an extension to the main optics box (see Figure~\ref{fig:snout}). After entering the dewar, the beam encounters a dichroic mirror mounted at a 45\degr\ angle which reflects the IR light into MIRSI and passes the optical light shortward of $\sim$1~\micron\ into the optical camera  through the exit window marked 3 in the figure  (see also the dichroic transmission/reflection curves in Appendix A). The exit window transmits in the optical and is opaque in the mid-IR, so it will emit thermal radiation into the region under the dichroic mirror. The dichroic does not transmit mid-infrared light, however, and it is cooled to the first stage radiation shield temperature ($\sim$47K), therefore the exit window does not contribute significantly to the background that the MIRSI detector sees, which is dominated by thermal emission from the telescope and sky. 

The original MIRSI window was made from KRS-5, which has transmission over the 0.6 - 30~\micron\ range. However, after some time of use at the IRTF, the window degraded and ``clouded'' over, having a frosted glass appearance. A spare window was substituted in, which in turn also degraded in the same way after a period of use. We attempted to have the
windows repolished, but they still had the clouded appearance, indicating it was not a surface effect. We decided to switch to a ZnSe window, which is a more durable material and has good transmission in the optical and 8-13~\micron\ window, although it cuts off near 20~\micron. This would enable MIRSI to do most of the science planned, and be more reliable until we could obtain a solution with KRS-5 that would not degrade.

In MIRSI, the telescope focus is at the aperture wheel positioned near the entrance, where an open aperture or slits can be selected. A fold mirror (M1) directs the light into the collimating mirror (M2) and then through the filters and pupil stop (see Figure~\ref{fig:optdesign}). The filter wheels and pupil stop are enclosed in a box that, along with other baffles, prevents stray light and out-of-band radiation from falling onto the camera mirrors M3 and M4 that reimage the focal plane onto the detector. The available filters are listed in Table~\ref{tab:EngPerf}, along with the measured sensitivity and instrument parameters used for each band. The transmission curves of the optical elements are shown in Appendix~\ref{sec:A1}, Figures~\ref{fig:ZnSeTrans} -- \ref{fig:filterC}. The use of cryogenic stepper motors that drive the motion via a gear on the circumference of the wheels eliminates the need for mechanical feedthroughs and thermal isolation of the motor shafts and ensures a reliable system which minimizes light leaks. 

MIRSI uses a Si:As blocked impurity band (BIB) detector array (320×240 pixels) developed by Raytheon \citep[Goleta, CA; see][]{1998SPIE.3354...99E}. The array is connected to a CMOS readout integrated circuit through indium bump bonds and mounted to a leadless chip carrier. MIRSI is configured to read out the detector in 16-channel mode, which corresponds to sets of 20 adjacent pixel columns on the array for each readout channel.

\subsection{Conversion to a cryocooler-based system}
The upgrade concept relied on the relatively simple design of the MIRSI dewar, where all of the critical components of the system are attached to the optical mounting plate, and contained within the main rectangular box of the dewar. This was easily separated from the cylindrical section containing the cryogen reservoirs and could be attached to a new upper section. The new upper section contains just the cryocooler cold head and a mounting plate that is supported by low thermal conductivity connections to the outer dewar shell. All of the optics, filters, motors, and detector array use their existing mounts and electrical connections to the outside of the dewar. Cryocooler vibration is a concern, but modern systems have lower vibration levels than previous models, and with careful isolation between the cold head and the instrument optics and detector, systems of this type have been demonstrated to have sufficiently low vibration so as to not degrade image quality \citep[e.g., ][]{2000Kataza,2010Jakob}. We have subsequently verified the low vibration by achieving the same image quality with the upgraded system compared to the liquid cryogen-cooled system (see \S3  below). The cryostat upgrade was performed by Infrared Laboratories\footnote{\url{https://www.irlabs.com/}} of Tucson, who built the original MIRSI dewar. The cold head used is a Sumitomo model RDK-415D2 two-stage Gifford-McMahon Refrigerator, with a model F-70L water-cooled helium compressor. 

\begin{figure}
    \centering
    \vskip 0.3in
    \includegraphics[width=3in]{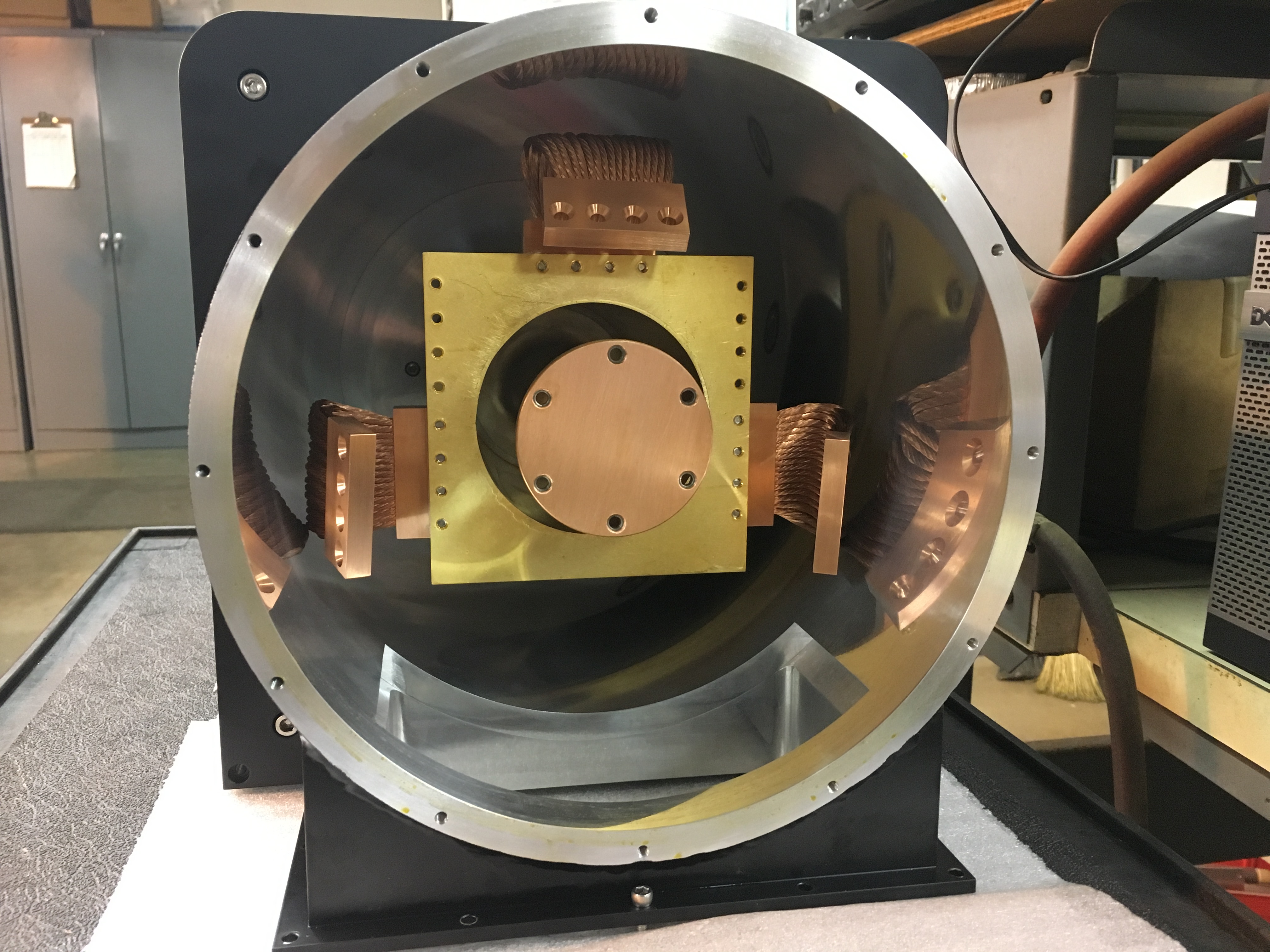}
    \caption{A view into the new upper section showing the attachment points of the stage 1 and stage 2 cryocooler. The three large braided copper straps are used to attach the outer shield to stage 1 of the cryocooler to isolate the instrument from vibration. A copper plate is bolted to the round attachment point of the stage 2 cooler in the center of the image, which extends down into the lower dewar where it uses similar copper straps to attach to the optical plate and inner radiation shield, and a separate path that goes directly to the detector stage. }
    \label{fig:isolate}
\end{figure}

\begin{figure}
    \centering
    \includegraphics[width = 0.49\textwidth]{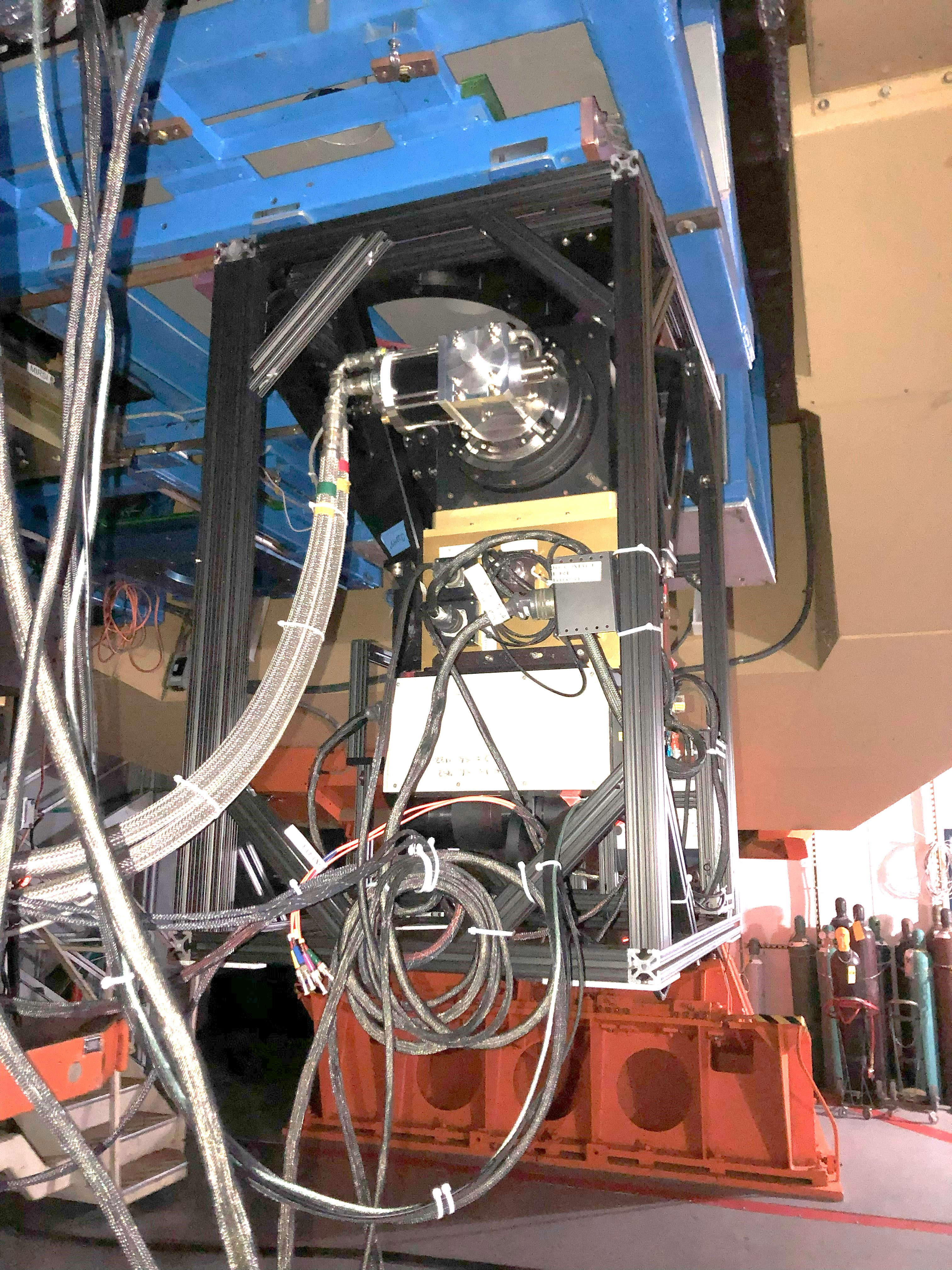}
    \caption{MIRSI mounted on the IRTF. In this photo, MIRSI is stowed in a position away from the telescope beam. The blue structure is the Multiple Instrument Mount under the telescope primary mirror. The gold MIRSI optics box is near the center of the image, with the cryocooler head and cooler lines above it. This view is rotated relative to the view in Figure~\ref{fig:CADmount}; here we are looking towards the side of the dewar with the electrical connectors which are seen on the right side of the optics box in Figure~\ref{fig:CADmount}.}
    \label{fig:OnIRTF}
\end{figure}

The cold head, which is black-anodized on the outside as depicted in Figures~\ref{fig:cryostages} -- \ref{fig:CADmount}, is mounted on a new upper section of the dewar that replaced the cryogen cans. Figure~\ref{fig:cryostages} shows the thermal design, with the figure on the left showing the parts of the dewar connected to the first stage of the cryocooler (the outer radiation shields), and the figure on the right showing the connection from the second stage to the detector and internal optics and radiation shields. Both of these stages are thermally connected via braided copper straps that act to mechanically isolate the optics and detector from the cryocooler head to minimize vibrations.  Figure~\ref{fig:isolate} shows a view into the new cold head section, showing the first stage connections to the copper straps. Figure~\ref{fig:OnIRTF} shows the full instrument mounted on the IRTF, with all of the compressor lines and electronics cabling attached.

To achieve the required array operating temperature, an isolated thermal connection was established that connects the second stage to the array mount (also via a braided copper strap) independent of the connection to the optics and inner radiation shields. The array is also thermally and electrically isolated from the optics plate using a G10 spacer. In order to electrically isolate the array and the array stage from the cryocooler and dewar case, a thin diamond sheet pressed between copper plates is used for high thermal conductivity. The stage can reach temperatures below 5~K, and can be thermally stabilized to within a few mK during operation using a heater resistor on the stage and an external Lakeshore controller to the nominal 6~K operating temperature. 

The system cooldown time is approximately 7 hours from room temperature until the first stage and outer shield reach equilibrium temperature ($\sim$ 47K for the outer shield), and a total of $\sim$13.5 hours for the detector stage and components on the optical plate to reach equilibrium (approximately 5K and 8K, respectively) with the detector powered off.

The use of the cryocooler as opposed to liquid cryogens has simplified operation at the telescope, but occasional maintenance is still required. Specifically, parts in the cold head will degrade over time and eventually it needs to be replaced. At the IRTF, this is done every two years, and has been done once since the upgraded MIRSI was delivered from IR Labs.

\subsection{Electronics and Mechanism Upgrades}
The MIRSI detector readout electronics were replaced by a system based on the same array controller and readout electronics used by other IRTF instruments. This solved the intermittent electronics issues present in the original MIRSI electronics, and makes it easier to maintain and repair when problems develop on the summit. The new electronics will be described in a separate paper.

The cryogenic stepper motor drivers were also replaced, including using Hall effect sensors instead of microswitches for improved reliability of wheel position sensing and accurate filter selection. The internal dewar wiring was also replaced due to the fragile nature of the original system.

\subsection{MIRSI Optical Camera (MOC)}\label{sec:MOC}
MIRSI is paired with a copy of the MIT Optical Rapid Imaging System \citep[MORIS;][]{2010DPS....42.4914G}, a fast readout optical camera used on the SpeX instrument at the IRTF. The new optical system, called the MIRSI Optical Camera (MOC), uses an Andor iXon 897 EM-CCD camera with a 512x512 detector and 16 micron pixels. A USB cable connects the detector electronics to a computer located on the bottom of the telescope that runs the software for controlling the camera. The detector is thermoelectrically cooled to -60 C to minimize dark current.  The field of view is 1 arcminute, and the pixel scale is 0\farcs12/pixel.  The camera optics consist of a field lens doublet, making a 5~mm diameter pupil image at a pupil stop, and a camera lens doublet creating an f/9.4 telecentric beam onto the detector.  The MOC has a filter wheel with SDSS r$'$, i$'$, and z$'$ filters, plus ND1, ND2, ND3, and ND4 neutral density filters to enable observing and guiding on optically bright targets.  The r$'$, i$'$, z$'$ magnitude zero points are 24.4, 24.1, and 23.5, respectively.

The optical beam passes through the dichroic which will contribute to some astigmatism in the MOC image, but the effect is minimal because of the slow telescope beam. The MOC optical design provides a spot size with 50\% encircled energy of 0\farcs12, and the typical optical seeing is in the range of 0\farcs6 - 1\arcsec, so the seeing dominates the image quality. The MOC and MIRSI mid-IR arrays have been placed as close as possible to being at the same focus position. Because the effects of atmospheric seeing are greater in the optical, we typically focus using the mid-IR image to achieve the best image quality.

The MOC has a user interface that allows one to either take single images or a series of consecutive images in guide mode. The user can set parameters such as the integration time, size of the guide window, guiding gain, and mode. The guider images can be saved if desired. A full description is given in the MOC user manual\footnote{\url{https://irtfweb.ifa.hawaii.edu/~moc/user/}}. When beamswitching to keep both A and B beam on the IR array, we typically guide in both beams as well. When guiding, the MOC software measures the source position relative to the center of the guiding box in the optical image, and sends commands to the telescope control system to center the source in the guide box at the frame rate of the guider images. After performing a beam switch or offset, the guider software calculates where the guide box should be on the optical array in the new position, and guides relative to that location (the ``commanded offset''). The commanded offsets are recorded in the IR image headers, so when reducing the data and constructing mosaics, the individual frames can be registered using the commanded offsets to high accuracy, even when the source is not visible in the IR images (see \S\ref{sec:guide}). 

The MOC enhances MIRSI’s capabilities by allowing the observer to place objects accurately on the IR field, and to guide on the optical emission from the science target or other nearby object in the field. This is especially important for spectroscopy since guiding errors result in the object moving off the slit, and without the capability to guide, multiple re-acquisition and/or offsets from nearby sources would be necessary. With the guide camera mounted to the same instrument, guiding errors due to different flexure of the science instrument and guide camera are minimized. The optical guiding is also important for observing faint targets which are not visible in individual MIRSI frames, as well as keeping a source centered on the slit in spectroscopic mode.

In addition to guiding, the MOC allows observers to perform simultaneous optical/IR photometry, a capability critical for thermal modeling of NEOs and other asteroids which requires near-simultaneous accurate optical and thermal fluxes to minimize uncertainties due to rotation and possible errors in the cataloged ``absolute magnitude" ($H$) values (the magnitude of an asteroid at zero phase angle and at unit heliocentric and geocentric distances\footnote{\url{https://cneos.jpl.nasa.gov/glossary/}}).

\begin{figure*}
    \begin{center}

    \includegraphics[height=2.5in]{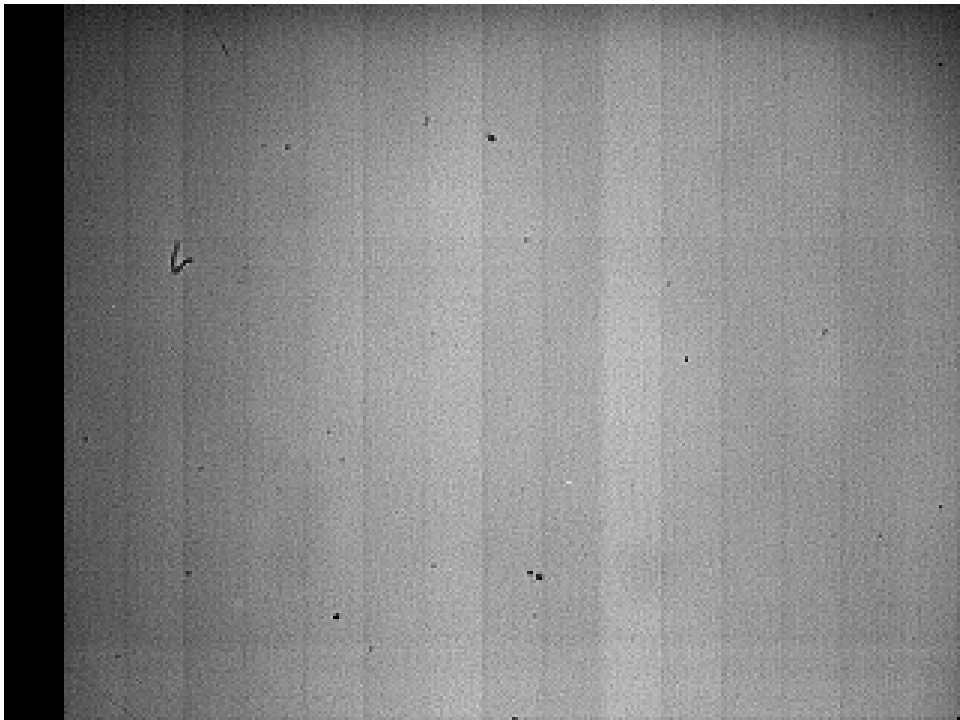}
    \includegraphics[height=2.5in]{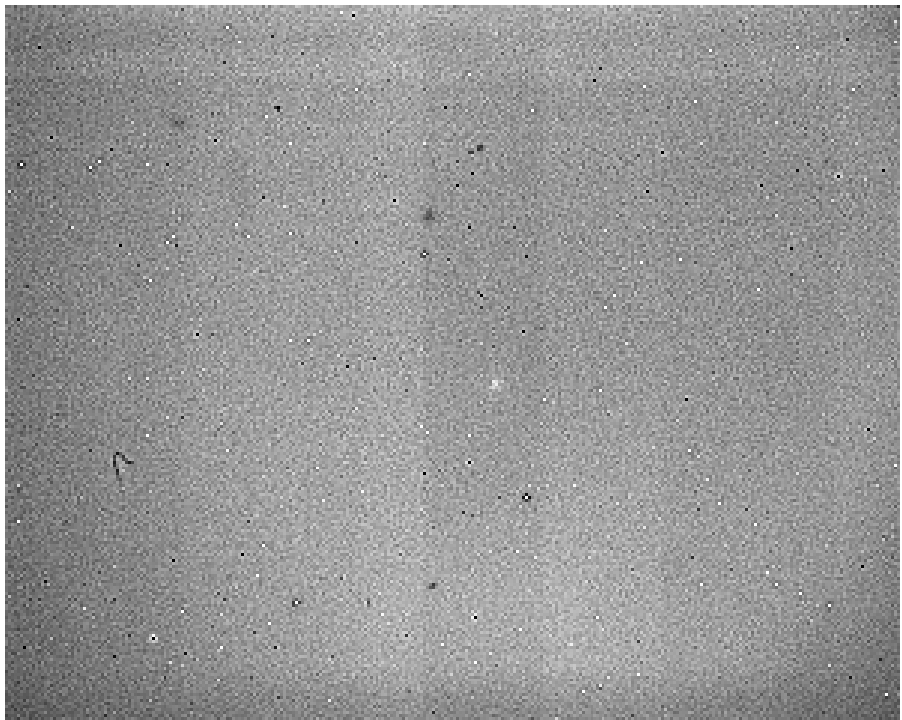}
    \includegraphics[height=2.48in]{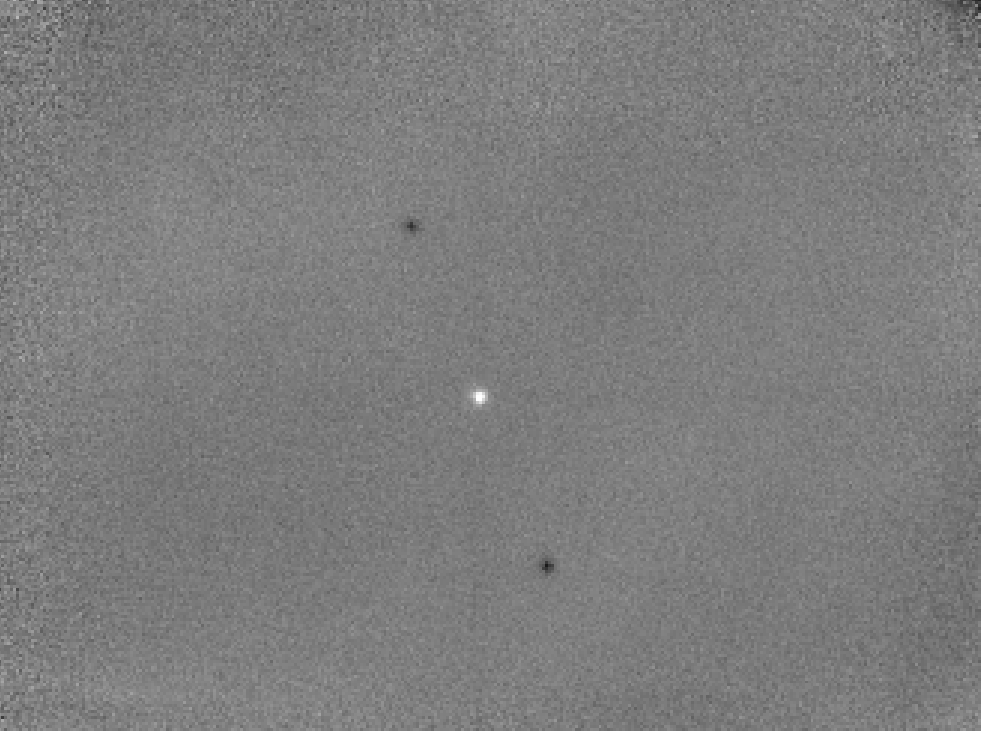}
    \includegraphics[height=2.52in]{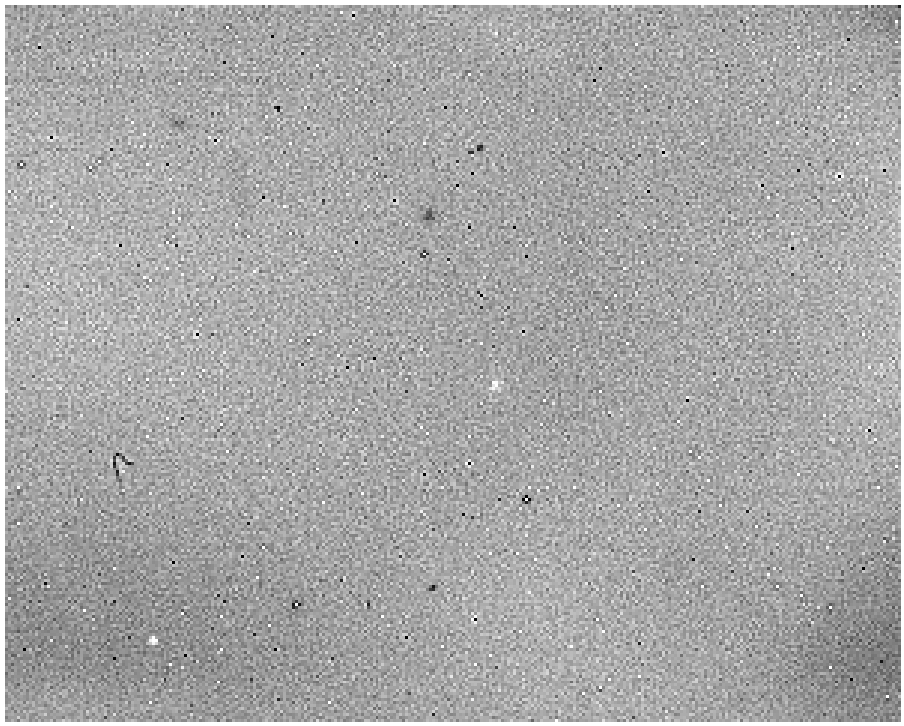}
    \end{center}
    \vskip -10pt
    \caption{ MIRSI 10.57~\micron\ images showing steps in the data reduction process for a NEO observation. The frames shown here were taken with the science-grade array currently installed in MIRSI. Top left: raw frame of 3.69 seconds integration time. On this night the leftmost readout channel was not operating, and appears black in this image.  In the raw frames written by the observing program, the image is oriented with North at the bottom and East to the left. Several bad pixel clumps appear dark in this image. Top right: the same frame after subtracting the subsequent nod frame and flipping the frame top-to-bottom  to orient the frame with North at the top and East to the left. In addition, the non-operational readout channel was trimmed. The positive and negative sources appear near the center of the image, with a beamswitch of 15\arcsec\ N, 6\arcsec\ W (the source was $\sim$2~Jy). Lower right: the subtracted image with column and row medians subtracted. Lower left: the final mosaic constructed from 54 individual 3.69 second integrations. The central bright source is the averaged source from both nod beams, aligned by guiding with MOC and using the commanded offsets for aligning the frames. The negative sources above and below are averaged between sky and off-source frames.}\label{fig:reduce}
\end{figure*}

\subsection{Observing and Data Reduction Techniques}

In ground-based mid-IR observing, one typically employs secondary chopping and telescope beamswitching to reduce sky noise and subtract the thermal background due to the sky and telescope. The minimum chop frequency required is in the range of 0.5 -- 1 Hz \citep[e.g.,][]{1983Papoular,1994Hoffmann,1999Miyata}. The chopping secondary is
no longer available at the IRTF, however they are working on a new secondary mount that will have chopping capability. Therefore, at the moment only nod and user offsets are possible for beamswitching. For point sources in uncrowded fields such as the NEO observations, we keep both beams on the array to maximize sensitivity and use both the A and B beam on-source data. The array is aligned with its long axis pointed E-W, and a typical nod throw is 6\arcsec\ W and 15\arcsec\ N. We also offset the telescope between nod pairs when taking a set of frames, so that we can correct for bad pixels and average out any residual pixel gain differences to produce the mosaics. 

Sample observations of a $\sim$2~Jy point source (an NEO) are shown in Figure~\ref{fig:reduce}. The upper left panel shows a raw image in the A beam, which is the coadd of 300 readouts using 0.0123 second frame time. During these observations the leftmost readout channel was non-operational, so it appears dark in this image. Each of the 16 channels reads 20 pixel columns of the array, which one can see in the image have slightly different offsets. The E-W component of the nod throw ensures  that the source falls on different output channels of the readout, so that any bright source artifacts that occur along the array columns are not present in both A and B beams.
The mean ADU level is $\sim$49,800 per frame. Some vignetting is visible in the upper part of the frame, highest in the upper right corner.

The upper right panel of Figure~\ref{fig:reduce} shows the result of subtracting consecutive A and B beam frames. The source is visible as a positive (white) source near the center, and a negative (dark) source above and to the left. Some bad pixel groups are visible at various locations around the image. The peak level of the source is $\sim$950~ADU in each frame. Offsets of $\sim$200~ADU are visible between the readout channels, and some horizontal stripes are seen near the top of the frame. The standard deviation of the pixels in the background is $\sim$300~ADU. In these observations the flux conversion factor was 1.10E-4 Jy/ADU.
The lower right panel of Figure~\ref{fig:reduce} shows the same image after subtracting column-wise and row-wise medians from the frame, which reduces the offsets between readout columns and the horizontal striping. 

The lower left panel shows the final mosaic made from 27 beamswitch pairs (54 frames total), cycling through 10 unique offset positions and using MOC guiding. The frames were aligned using the commanded offsets and beamswitch 
vector. Both the positive and negative images of the nod pair are used when making the mosaic: each difference image is multiplied by -1 to make a second image with the negative source (from the B beam position) positive. Those images are then shifted to align them with the A-beam source positions, and the images are averaged with sigma clipping. This results in a final mosaic that has a central positive source that combines all beams, and negative residuals showing up both above left and below right of the source which are the combination of blank sky and the negative source from half of the frames. Those residuals are ignored and the photometry is performed on the central positive source.

We have a basic reduction pipeline for the MIRSI IR data that consists of three python programs. The first program reads in the raw frames, performs the A-B subtraction, column and row median corrections, and writes the difference images similar to the lower right panel of Figure~\ref{fig:reduce}. The second program reads in these difference frames and constructs a mosaic such as in the lower left panel of Figure~\ref{fig:reduce}. There are several options for how to perform the relative shifts between frames to construct the mosaics: one can either use the commanded offsets if guiding with MOC, determine the offsets by cross-correlating on the individual frames (requires that the source is bright enough to be visible in each frame), or have the user interactively choose the location in each frame to center on. The third program performs aperture photometry on the frames or mosaics. The pipeline is included with the supplementary materials of this paper, and the repository for the current version of the pipeline is on github\footnote{\url{https://github.com/jhora99/MIRSI}}.

\section{Performance with Engineering Grade Array: 2021 - 2023}
We had available to us an "engineering grade" array from the original MIRSI development that we first installed in the upgraded MIRSI for testing and initial commissioning of the instrument. This array has some cosmetic and uniformity issues that caused it to be classified as engineering grade, but otherwise has a similar sensitivity to the science-grade detector.
The upgraded MIRSI was made available to general observers on the IRTF for the 2022A-2023A semesters, where it was scheduled for use for parts of 25-28 nights per semester. Unfortunately, MIRSI's current sensitivity is significantly worse than the original MIRSI as detailed below, approximately a factor of 10 less sensitive in all bands. However, it was still possible to execute several science programs, including observations of planets and asteroids. Some examples are given in the following sections.

\subsection{IR Sensitivity}
The sensitivity in each filter measured with the engineering-grade array is shown in Table \ref{tab:EngPerf}. The values in the table were calculated based on observations of $\alpha$~Tau obtained on the nights of 2021/10/01 and 2021/10/02 at an airmass of $\sim$1.02, so no airmass correction was done. The ITIME is the on-chip integration time, and the 1$\sigma$ sensitivity in 10 minutes is based on the actual elapsed time spent observing the star, including overheads for telescope nodding and array readouts. The measurement in each filter used 20 beamswitched/dithered frames. The beamswitch and dithering offsets were small enough to keep the star on the array at all times, to maximize the on-source time and sensitivity of the observations. The point source sensitivity numbers are based on the per pixel noise and the equivalent noise area \citep{1983King} for each wavelength, assuming the diffraction-limited PSF size and 0\farcs5 seeing.

\subsection{Guiding Performance with MOC}\label{sec:guide}
The performance and operation of the MOC is similar to that of the MORIS system. We have demonstrated that the MOC can successfully guide on point sources as faint as $V$-band magnitude (Vega) $\sim$17 with exposure times of a few seconds. Typical dither dwell times for the IR observations are on the order of 20 seconds, so this allows for several corrections at each position for faint sources. For brighter sources, one can use integration times of 1 second or less, although the most common pointing error is a slow drift that would move the telescope by a fraction of an arcsec in a minute or more of time.

\begin{figure}
    \centering
    \includegraphics[width=0.222\textwidth]{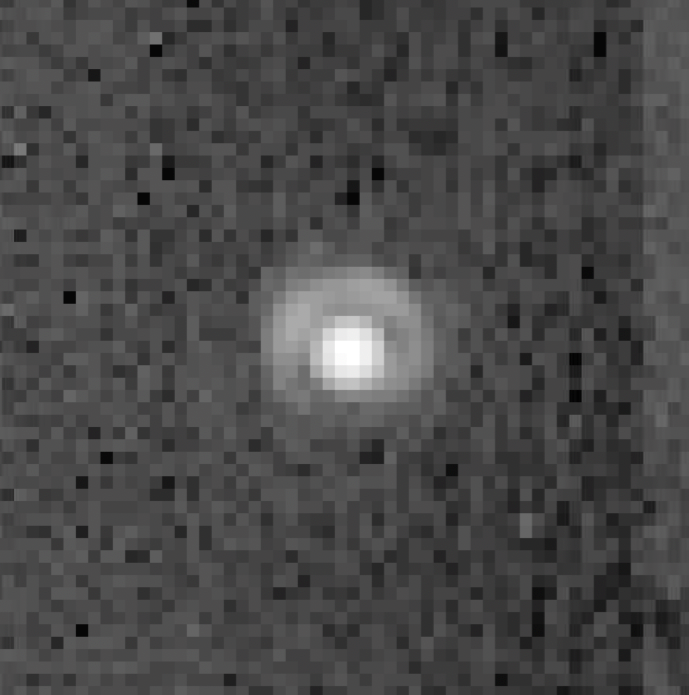}
    \includegraphics[width=0.222\textwidth]{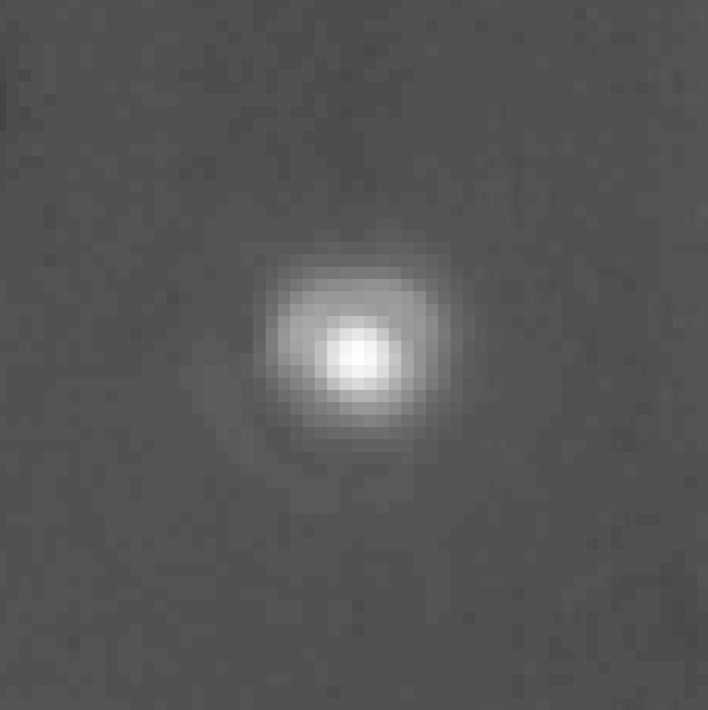}
    \includegraphics[width=0.45\textwidth]{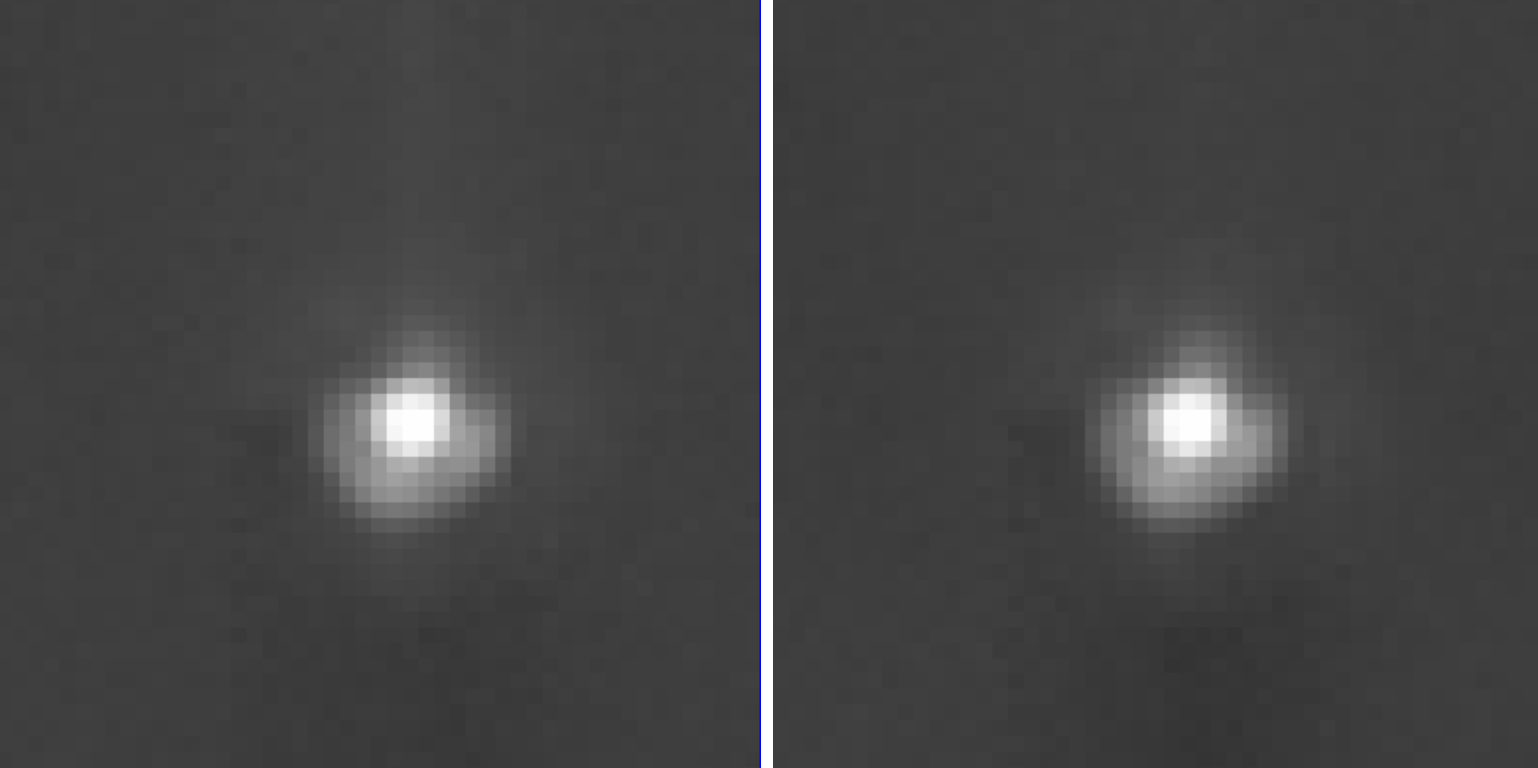}
    \caption{Images of standard stars taken with MIRSI and reduced using the MIRSI pipeline.  Top Left: single background-subtracted frame of $\alpha$~Tau at 11.7~\micron, taken with a total exposure time of 0.75 seconds. The Airy disk and first ring of the diffraction pattern of the telescope are visible. Top Right: averaged stack of 10 dithered frames of $\alpha$~Tau at 11.7~\micron~while guiding with MOC, and aligned using the commanded offsets. The FWHM of the single frame is 1\farcs02, the FWHM of the ``blind"-stacked guided frames is 1\farcs08 (expected diffraction-limited FWHM is $\sim$0\farcs83). 
    Bottom Left: image of HR1457 obtained on 2021/10/02 at 10.57~\micron~ while guiding with MOC, and the images aligned using the commanded offsets. Bottom Right: image of HR1457 using the same data as the image at the bottom left, except aligning the frames using the centroid of the source in each frame. The images have the same FWHM of 0\farcs81, compared to the expected diffraction-limited FWHM of 0\farcs75.}
    \label{fig:guiding}
\end{figure}

Figure~\ref{fig:guiding} shows the accuracy of how images can be aligned with guiding and blind stacking of the images. The images in the top row show a single image of $\alpha$~Tau and a mosaic image where the frames have been shifted and coadded according to the telescope offsets. The FWHM of the images are nearly identical. One does not normally need to guide for bright calibration stars since one can align the images based on the IR frames themselves, but this mode is especially important for cases where the target of interest is not visible in the individual IR frames, such as the NEO project discussed in \S \ref{section:NEO} below.
In the bottom row, a similar set is shown for the star HR1457 obtained under better seeing conditions. The mosaic calculated from aligning the individual frames based on the source centroid is almost identical to the mosaic using the commanded offsets when guiding with MOC.

\begin{deluxetable*}{rrrrrrr}
\tablecaption{MIRSI Sensitivity (Engineering Array, October 2021)}\label{tab:EngPerf}
\tablewidth{0pt}
\tablehead{
\colhead{} & \colhead{Assumed}&\colhead{}& \colhead{} & \colhead{} & \colhead{per pixel} & \colhead{point source}\\ 
\colhead{} & \colhead{$\alpha$ Tau Flux}&\colhead{} & \colhead{}& \colhead{ITIME} & \colhead{1$\sigma$, 10 min} & \colhead{1$\sigma$, 10 min}\\
\colhead{Filter} & \colhead{(Jy)}&\colhead{Jy/ADU} & \colhead{Coadds} & \colhead{(sec)} & \colhead{(mJy)} & \colhead{(mJy)} 
}
\startdata
2.2	& 8146 & 0.63925&	100&	0.005&	42.1&	214\\
4.9	& 2081 & 0.19871&	100&	0.005&	20.6&	107\\
7.7	& 942 & 0.20587&	50&	0.007	&89.2	&556\\
8.7	& 763 & 0.1381	&50	&0.005	&33.2&	222\\
9.8	& 647 & 0.09425&	50	&0.007&	31.0	&224\\
10.57 & 555 &	0.02542&	100	&0.005	&4.6	&35\\
11.7 & 481 &	0.03363&	50	&0.015	&10.1	&84\\
12.28 & 438 &	0.26359&	100	&0.01	&28.2	&242\\
12.5 & 424 &	0.0352&	50	&0.015&	16.2&	141\\
Q0 & 254 &	0.87503	&500&	0.06&	473.0&	5310\\
Q1 & 238 & 0.64478	&500&	0.06&	376.0&	4385\\
Q2&	229 & 0.88396	&500&	0.06&	558.9&	6637\\
20.7 & 180 	&0.77818&	200	&0.005&	98.4&	1331\\
\enddata
\end{deluxetable*}

\begin{figure}[h]
    \centering
    \includegraphics[width=3.3in]{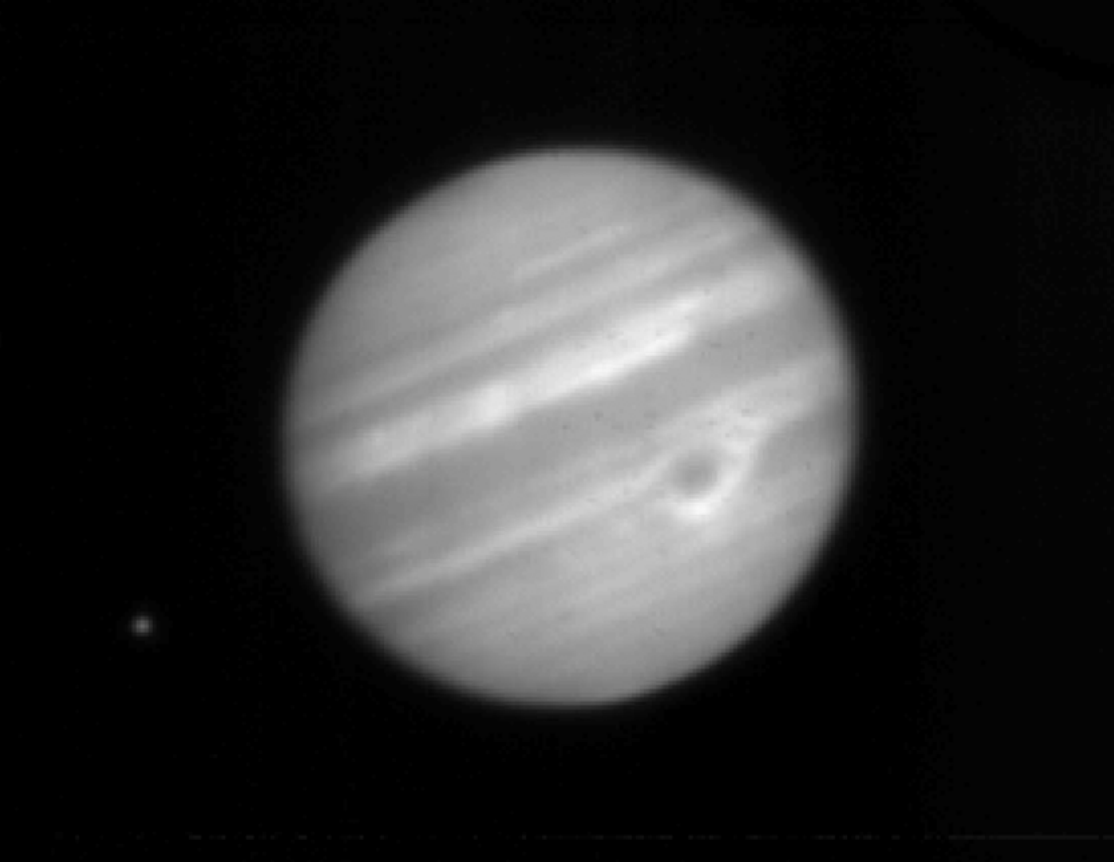}
    \caption{A MIRSI N-band image of Jupiter and Io, obtained in the fall of 2021. The Great Red Spot is visible in the lower right part of Jupiter.}
    \label{fig:jupiter}
\end{figure}

\begin{figure}
    \centering
    \includegraphics[width=3.2in]{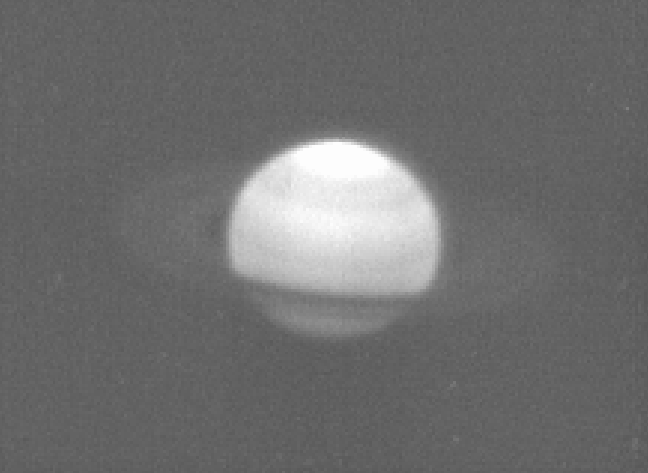}
    \caption{MIRSI N-band image of Saturn, taken in the fall of 2021. Compare to Figure 1 of  \cite{2017A&A...599A..29F} that shows multiwavelength mid-IR images obtained with the COMICS instrument on Subaru. }
    \label{fig:saturn}
\end{figure}
\subsection{Examples of Astronomical Results}
\subsubsection{Planets}
We show sample images taken with the MIRSI engineering grade array in Figures \ref{fig:jupiter} -- \ref{fig:saturn}. These were taken with the N-band filter and are mosaics composed of several different dither positions. In the case of Jupiter, the object was moved off the array with beamswitch commands to obtain the sky reference images. For Saturn, the object was moved to a different position on the array in order to minimize the time needed to reach the signal-to-noise necessary to detect the ring emission. 

\subsubsection{NEOs}\label{section:NEO}
\cite{2022DPS....5451405L} began a program of NEO observations with the engineering array version of MIRSI in 2021, following the goals of the MIRSI upgrade proposal described in \S \ref{MIRSIupgrade}. In our simultaneous \citet{Lopez-Oquendo2024} work, we present a detailed description of the MIRSI-NEO program along with initial results from the 2022-2024 survey. See Figure~\ref{fig:reduce} for an example of an NEO observation. We also utilized MIRSI as part of the International Asteroid Warning Network rapid response characterization campaign focused on the newly discovered NEO 2023 DZ2 \citep[][]{Reddy2023, Reddy2024} to estimate the object's diameter and albedo. This activity demonstrated the utility of MIRSI at the IRTF to characterize a newly discovered asteroid on short notice to assess its potential impact threat.

\subsubsection{Star Formation Regions}
\begin{figure}[h]
    \centering
    \includegraphics[width=3.3in]{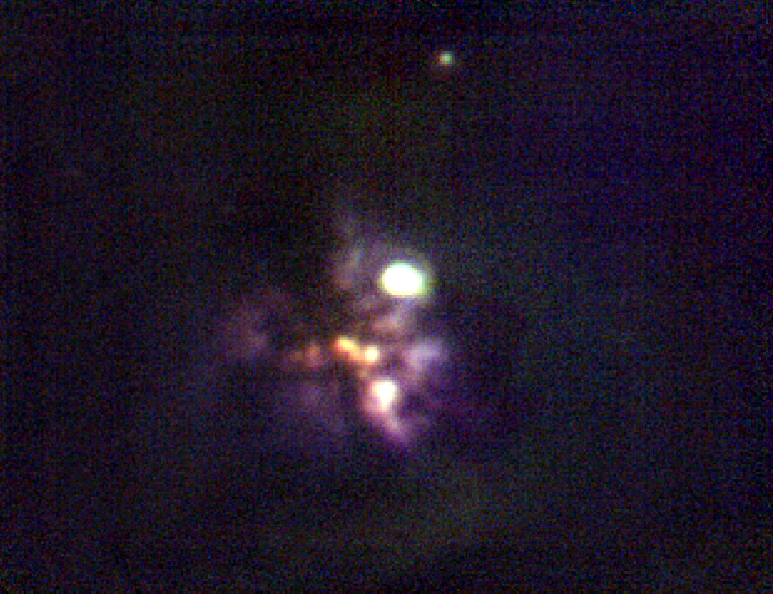}
    \caption{MIRSI color image of the BN/KL object in Orion. Blue is 8.7~\micron, green is 11.7~\micron, and red is 12.5~\micron. The image is $\sim$1\farcm35 wide. Compare to previous mid-IR imaging of this region from the IRTF by \citet{Gezari_1998} and \citet{2006ApJ...637..823K}.}
    \label{fig:bnkl}
\end{figure}

Figure~\ref{fig:bnkl} shows a color image of the BN/KL region in Orion at 8.7, 11.7, and 12.5~\micron, obtained on 2021/10/01 using the engineering-grade array. The mosaic at each wavelength combined 20 dithered/beamswitched frames taken with ITIME=0.015~s and 200 coadds. The frames were aligned according to the peak emission point in the image. The color scaling is set to enhance the lower-level emission, so the bright BN object is saturated in all colors and appears white in the figure.

\subsection{Science-Grade Array}
MIRSI was removed from the telescope in the fall of 2023 to switch detector arrays. A science-grade array that had been used in a now-decommissioned instrument was loaned to the IRTF for use in MIRSI. The new array has been used since the start of the 2024A semester. The array has fewer bad pixels and has better uniformity of pixel response (see Figure~\ref{fig:reduce}), but the overall sensitivity is not significantly changed. This indicates that some other change in the system has taken place compared to the original MIRSI system, for example a degradation of optical component(s), an 
alignment issue, or a non-optimal detector readout scheme. These possibilities are currently under investigation. 

\section{Summary and Future Work}
MIRSI's cryogenic system has been upgraded and the instrument is back in operation at the IRTF, available to observers since the spring 2022 semester. The instrument is mounted on the telescope Multiple Instrument Mount along with the other facility instruments and can be swapped in quickly as needed. The cryocooler system keeps MIRSI at operating temperature with much reduced operational cost and complexity compared to the liquid cryogen-based system.

The MIRSI instrument control program has been replaced by an IRTF-standard graphical user interface that will seem very familiar to users of other IRTF facility instruments. The three mechanisms that control the aperture wheel and the two filter wheels are operated with simple drop-down menus. The frame time, number of cycles and coadds, and beam pattern are controlled from the main menu. Sequences of observations can be programmed using macros, which can also configure the instrument and set up all of the observing parameters. 

The addition of the MOC has made possible many programs where IR sources are not available to guide or align frames. This has been used extensively in the NEO observing program, but any application where optical guide objects are available but the IR source is too faint or extended to guide on will benefit from this capability. The relative flexure between MOC and MIRSI is sufficiently low that one can use the commanded offset positions to align frames with little effect on the image quality of the IR mosaics.

As previously mentioned, a chopping secondary is planned for the IRTF, which will improve sensitivity by reducing the noise from fluctuations in the sky background. Currently we are trying to determine the reason(s) behind the lower sensitivity the current MIRSI system has compared to its original performance \citep{2008PASP..120.1271K}. Some of the lower sensitivity
is due to higher noise because we are not using a chopping secondary,
but we estimate this is a factor of $\sim$2 effect and does not account
for all of the reduced sensitivity. Updates to MIRSI status and sensitivity are posted to the IRTF web site\footnote{\url{https://irtfweb.ifa.hawaii.edu/
}} before proposal submission deadlines each semester.

\begin{acknowledgements}
We acknowledge the significant cultural role and reverence
that the summit of Maunakea has within the indigenous Hawaiian
community and that we are most fortunate to have the opportunity
to conduct observations from this mountain. The Infrared Telescope Facility is operated by the University of Hawaii under contract 80HQTR24DA010 with the National Aeronautics and Space Administration.
This work was partially funded by a grant from the NASA Solar System Observations/NEOO program (NNX15AF81G). The original MIRSI instrument was funded by NSF grant 9876656 and support from Boston University.
    \software{Astropy \citep{astropy},
Matplotlib \citep{matplotlib}}
\end{acknowledgements}

\appendix
\section{Transmission of Optical Elements}\label{sec:A1}
\begin{figure}[h]
    \centering
    \includegraphics[width=0.49\linewidth]{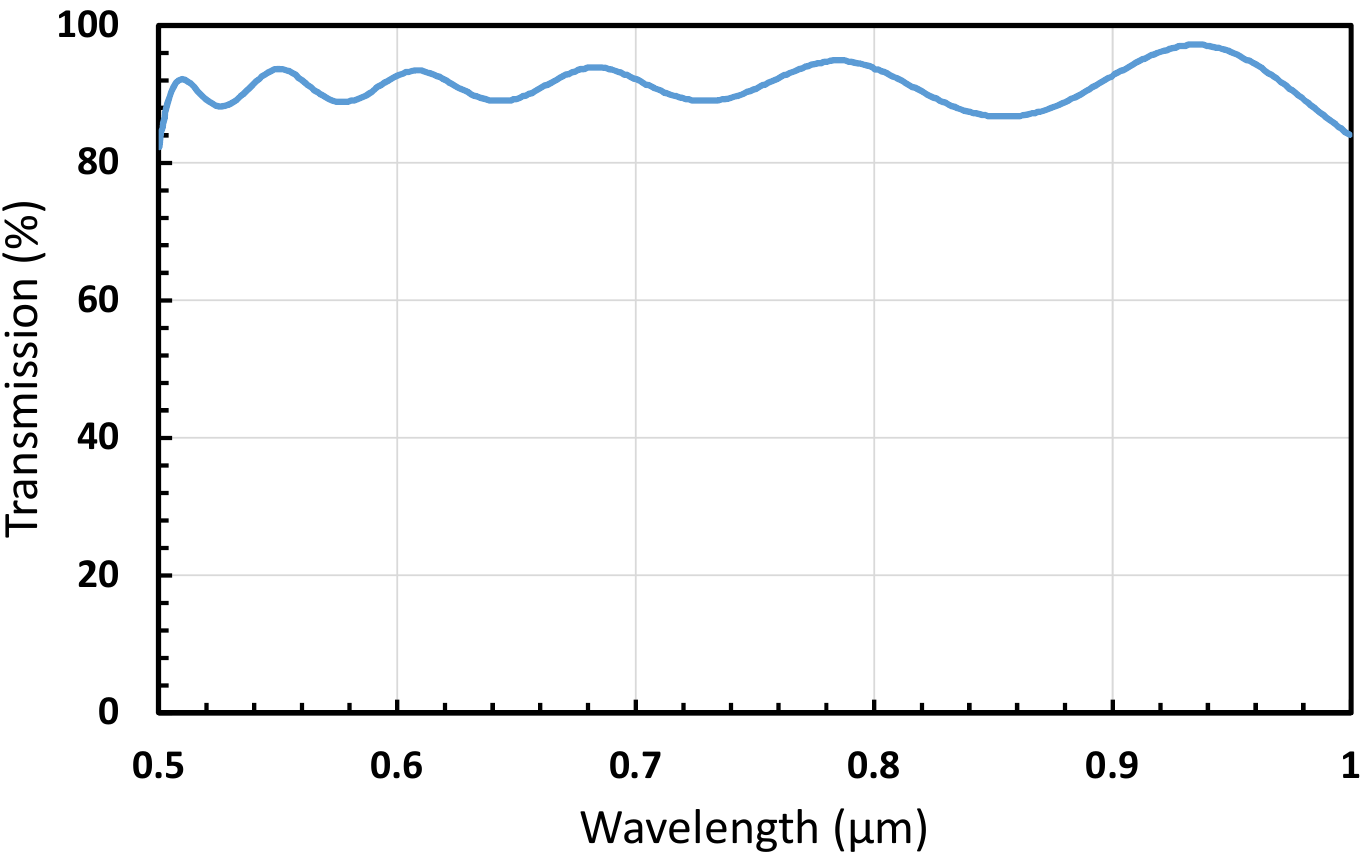}
    \includegraphics[width=0.49\linewidth]{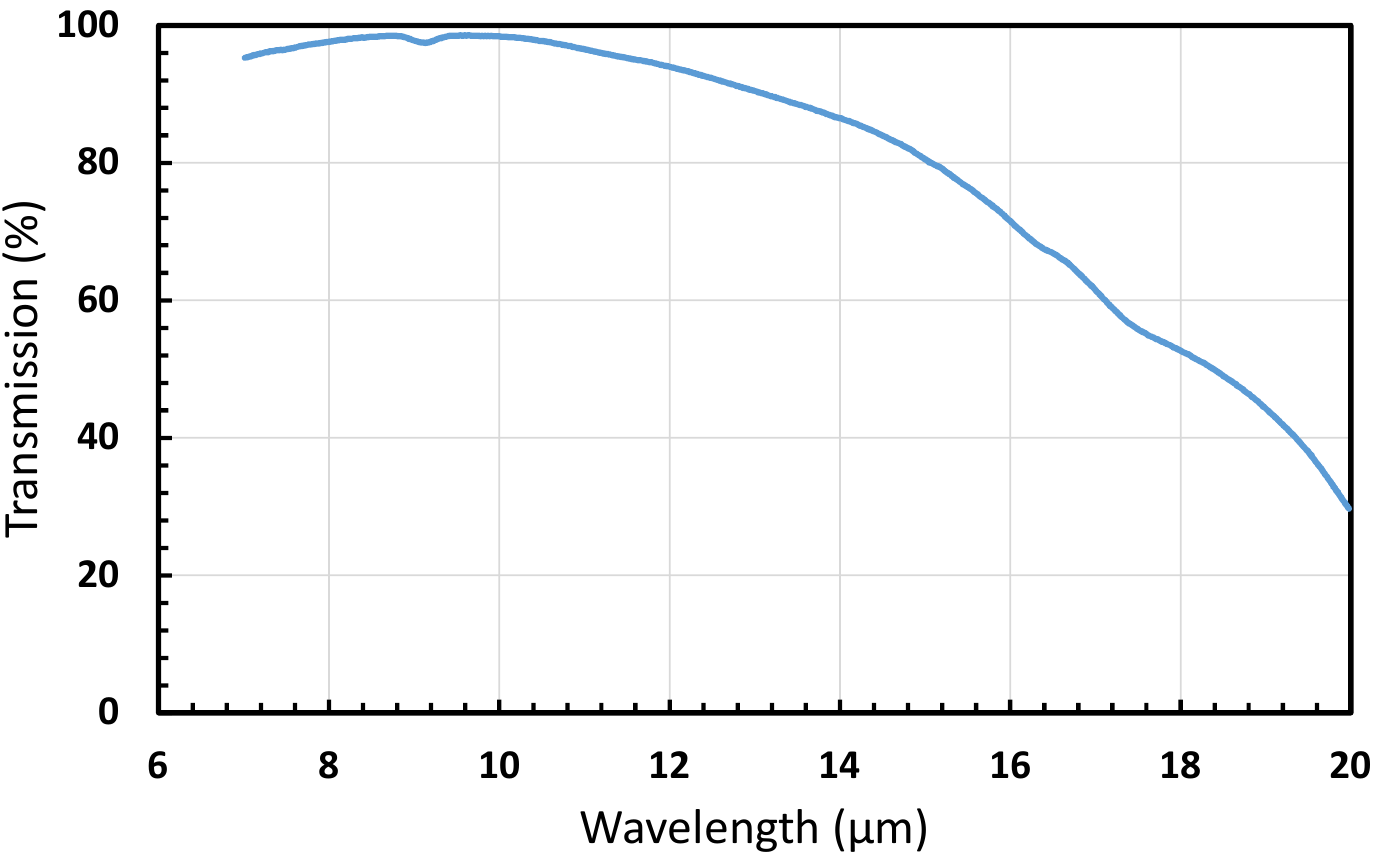}
    \caption{Manufacturer-measured transmission curves for the antireflection-coated ZnSe dewar window. Left: the optical wavelength range (0.5 - 1 \micron). Right: The mid-IR range (7 - 20 \micron).}
    \label{fig:ZnSeTrans}
\end{figure}

\begin{figure}[h]
    \centering
    \includegraphics[width=0.49\linewidth]{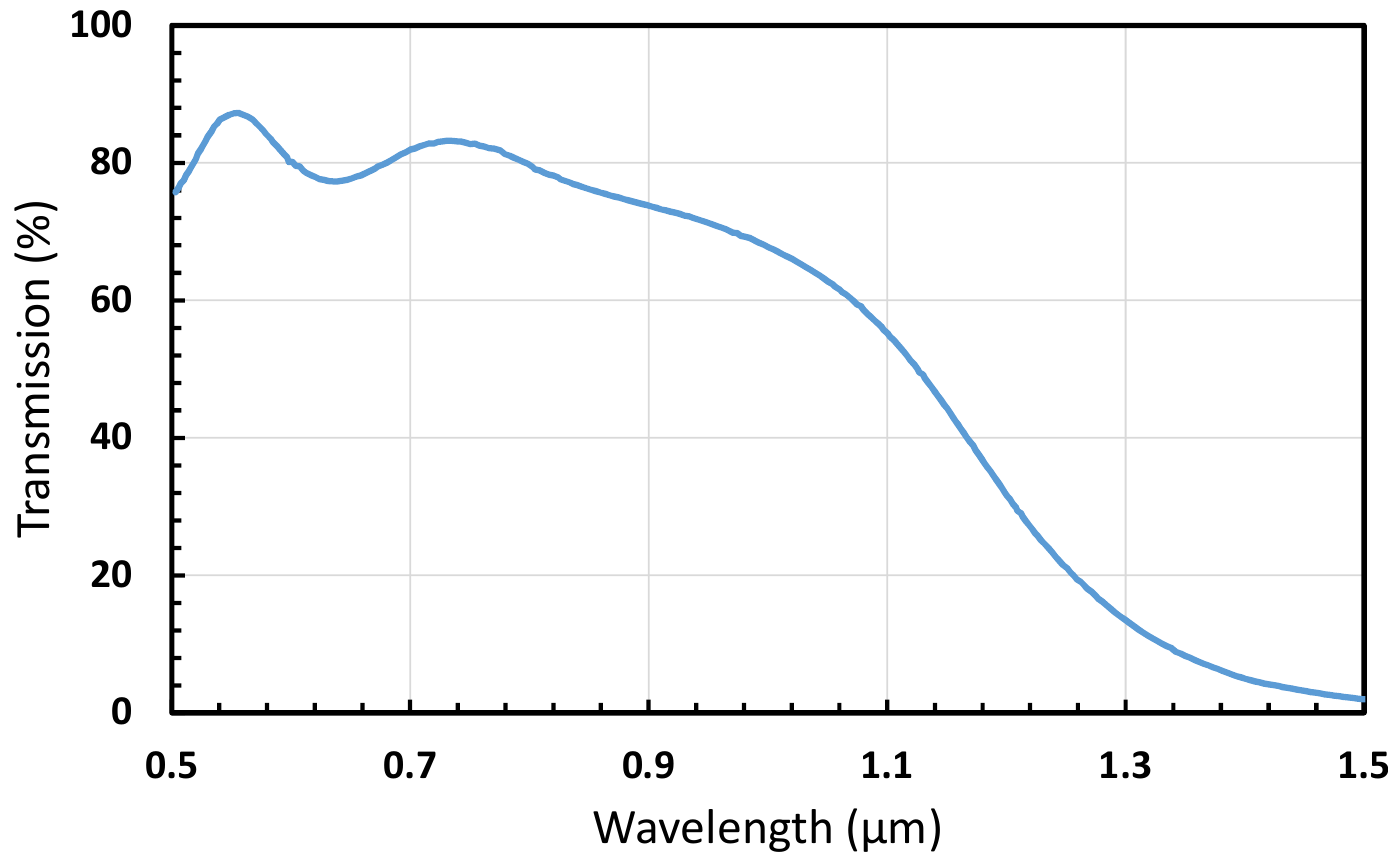}
    \includegraphics[width=0.49\linewidth]{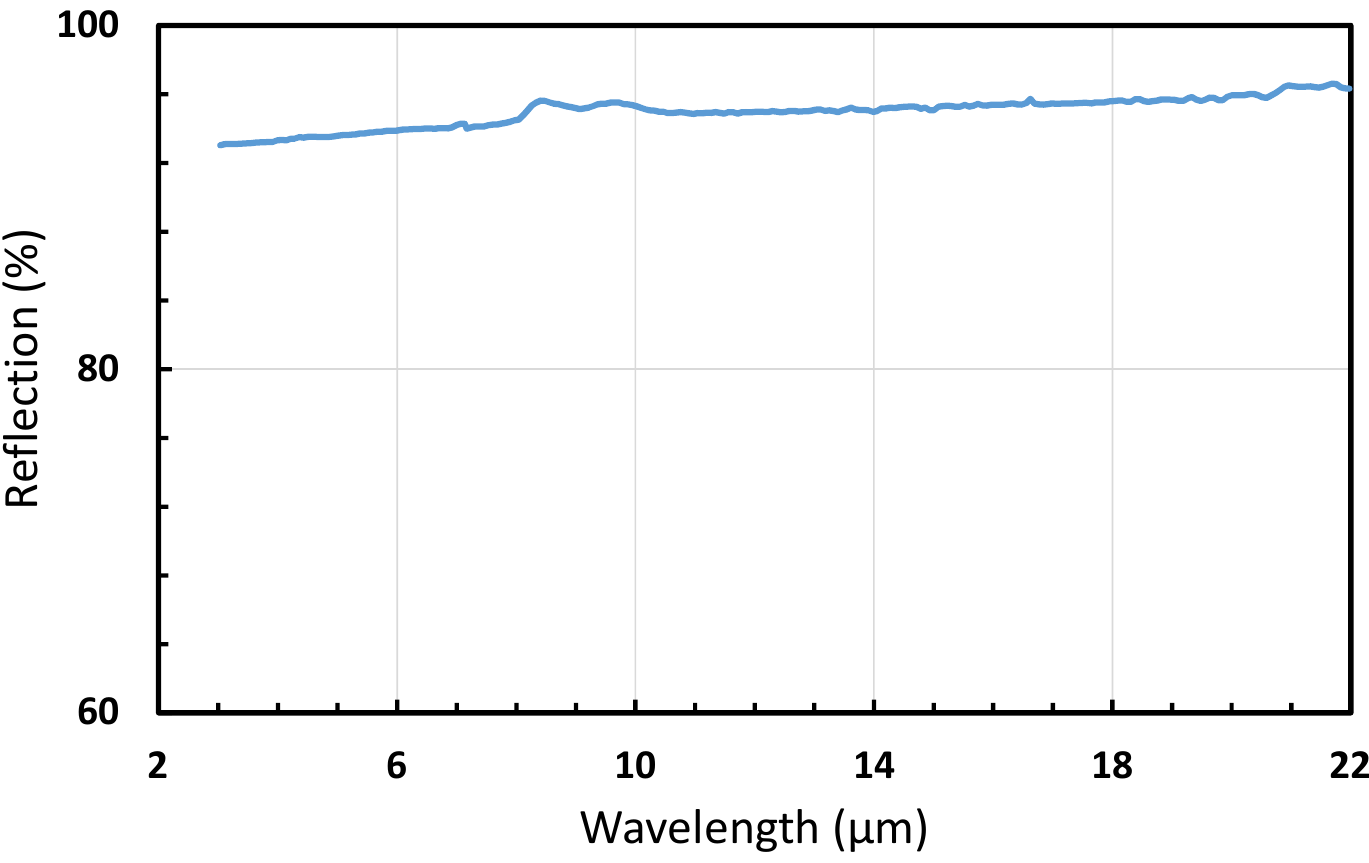}
    \caption{Manufacturer-measured curves for the dichroic. Left: The transmission curve (optical light transmitted to the MOC), measured at 0\degr\ angle of incidence. Right: the reflection curve for light going to the IR optics, measured at 12.5\degr\ angle of incidence.}
    \label{fig:Dichroic}
\end{figure}

\begin{figure}
    \centering
    \includegraphics[width=6in]{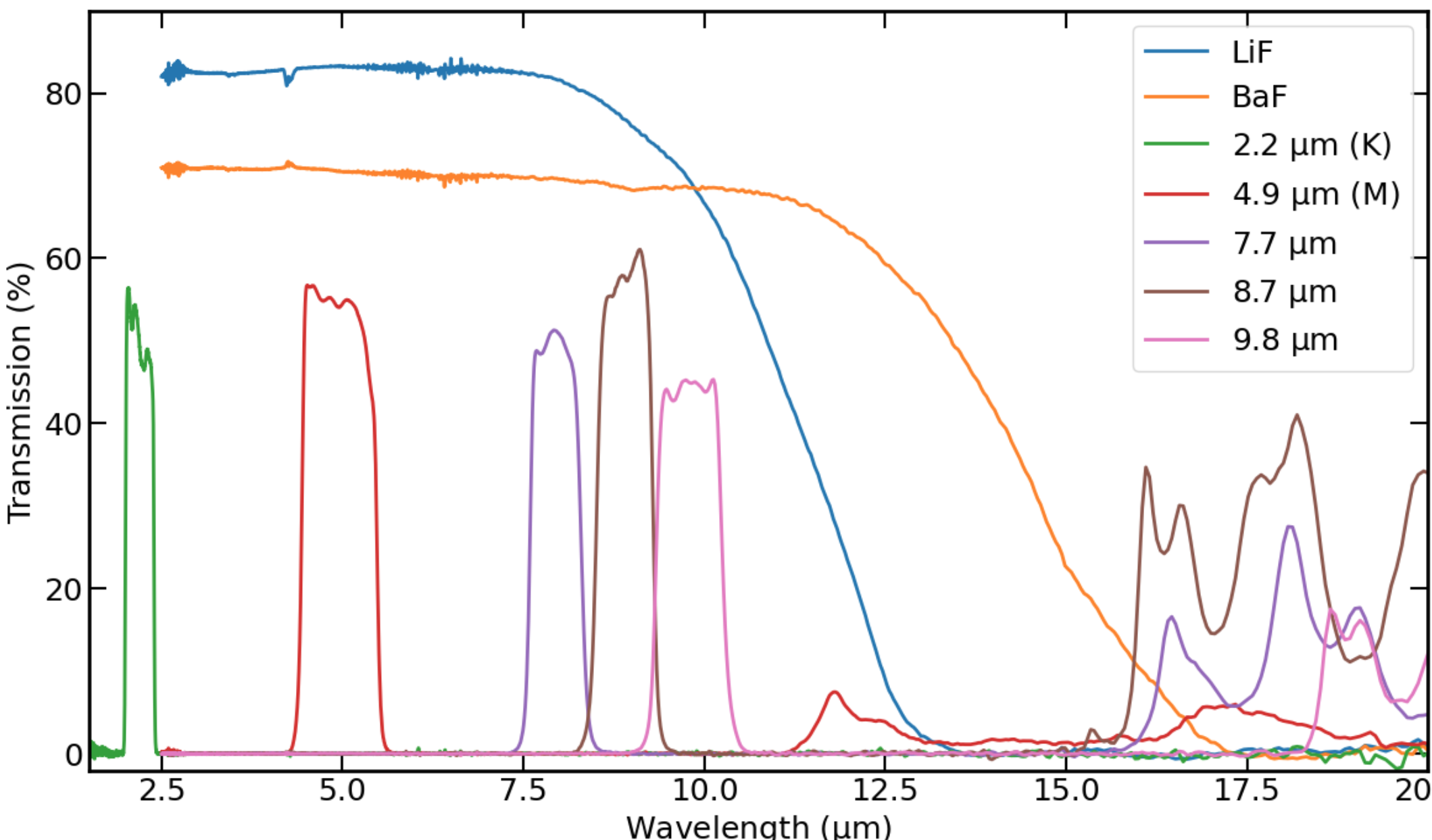}
    \caption{Scans of the MIRSI filters performed at the IRTF at room temperature. At operating temperature, the transmission curves will have close to the same shape but will shift to shorter wavelengths by $\sim$2\%. The traces are labeled with the MIRSI filter names. The LiF blocker is used in series with the 4.9, 7.7, and 8.7~\micron\ filters and the BaF blocker with the 9.8~\micron\ filter to suppress the long-wavelength transmission beyond 11~\micron.}
    \label{fig:filterA}
\end{figure}

\begin{figure}
    \centering
    \includegraphics[width=6in]{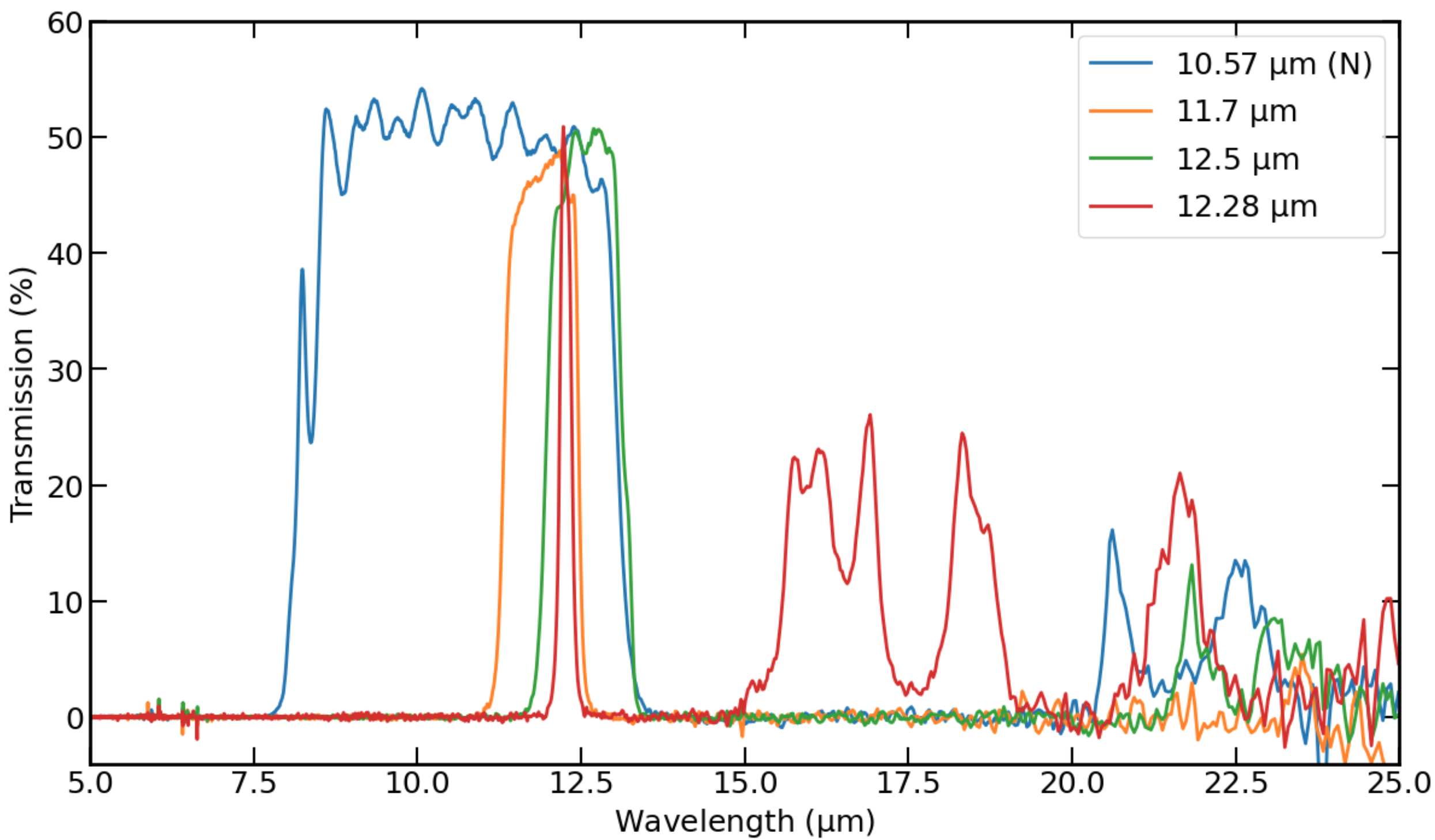}
    \caption{Similar to Figure~\ref{fig:filterA}, except for the MIRSI filters as noted in the legend. The BaF blocker is used in series with the 11.7, 12.5, and 12.28~\micron\ filters to suppress their long-wavelength transmission.}
    \label{fig:filterb}
\end{figure}

\begin{figure}
    \centering
    \includegraphics[width=6in]{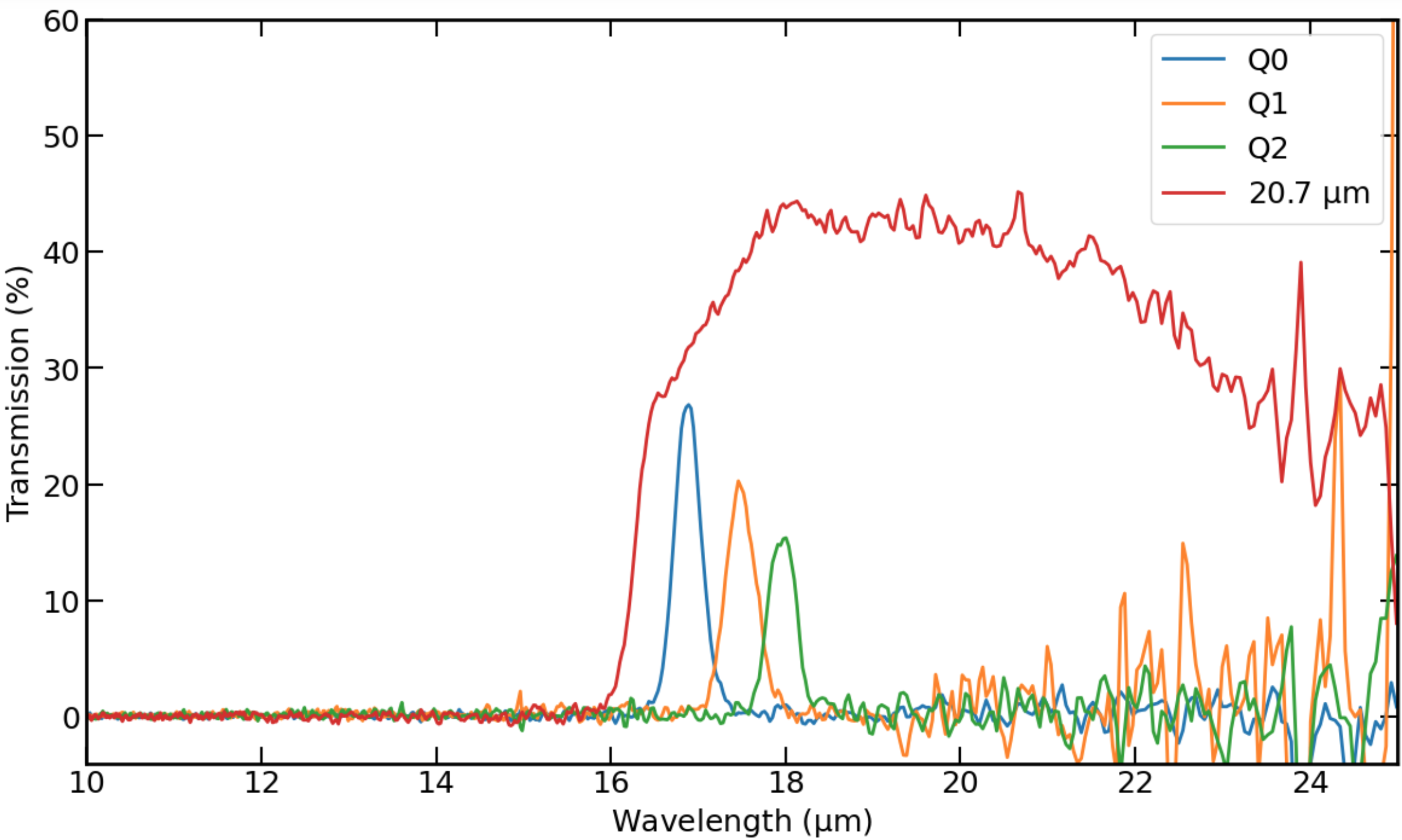}
    \caption{Similar to Figure~\ref{fig:filterA}, except showing the Q-band filters. Note that the window transmission will also significantly affect the transmission through the system at these wavelengths (see Figure~\ref{fig:ZnSeTrans}).}
    \label{fig:filterC}
\end{figure}

Plots of the transmission and reflection of the various optical elements are shown in Figures \ref{fig:ZnSeTrans} -- \ref{fig:filterC}. The dewar window and dichroic scans were provided by the manufacturers. The MIRSI filter transmission curves were also measured at the IRTF at room temperature using a Thermo Nicolet FTIR spectrometer.

The mirrors in the MIRSI optics are all gold-coated, the flat mirror is fused silica and the elements with power (M2-M4) are diamond-turned aluminum. Each of these elements have reflectivities of $\geq$98\% over MIRSI's operating range.

The transmission and reflectance data plotted in these figures are available as ASCII-format files to download from the electronic version of this paper.


\begin{figure*}
    \centering
    \includegraphics[width=0.99\linewidth]{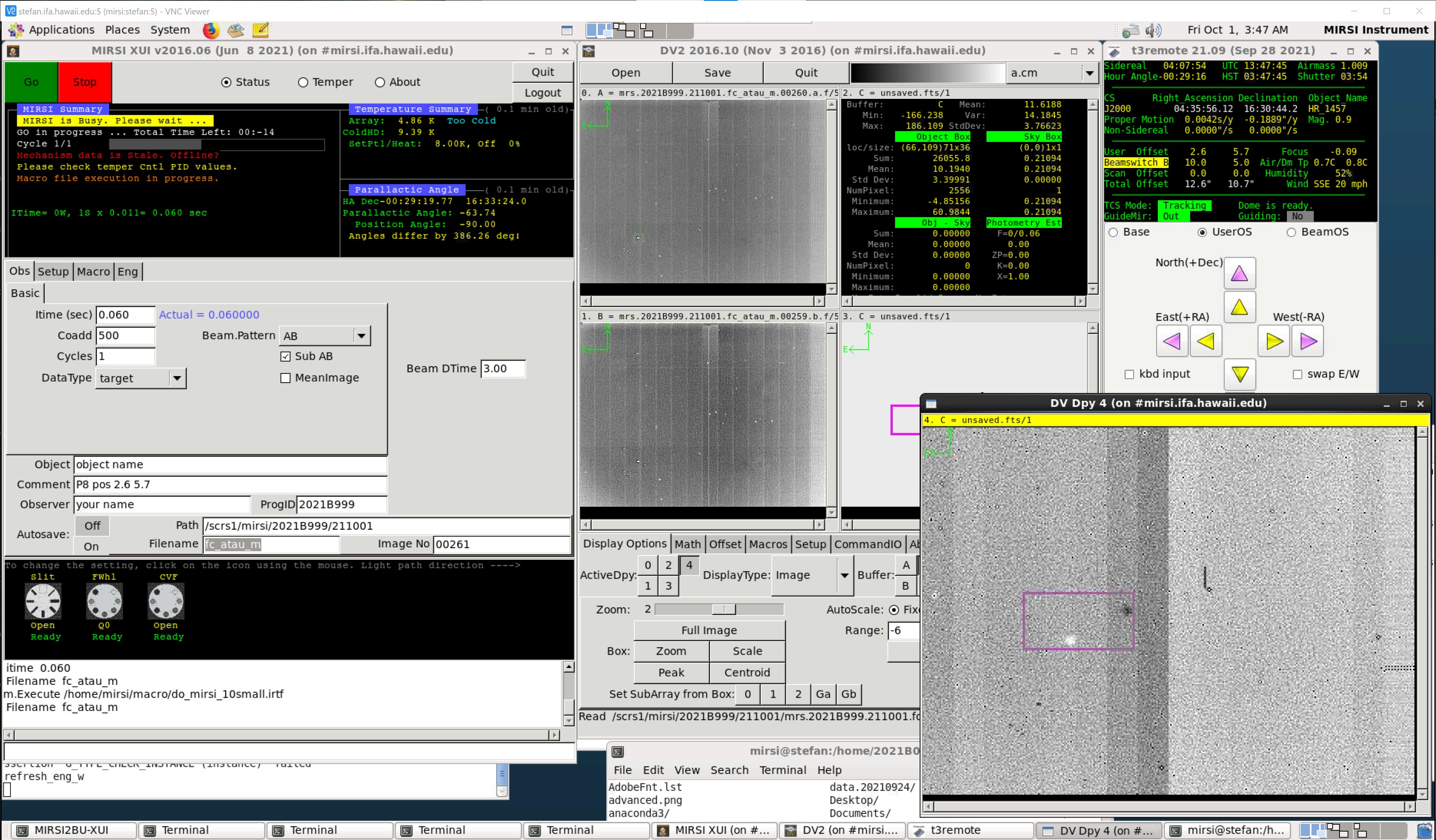}
    \caption{Screenshot of the MIRSI GUI, showing observations of the star $\alpha$~Tau in the Q0 filter. Observations are being taken in the AB beam pattern, and the difference image is being displayed in image buffer 4 which is the large image in the lower right corner. Offsets between the detector readout channels are seen as vertical stripes in the background (20 columns per channel), and dead pixels and other array artifacts are visible in the image.}
    \label{fig:gui}
\end{figure*}

\section{MIRSI Graphical User Interface (GUI)}
A screenshot of the MIRSI GUI is shown in Figure~\ref{fig:gui}. It is similar to the other IRTF instrument control programs, which makes it easier for experienced IRTF users to use and for the telescope staff to maintain. The user can change the camera parameters of Itime (on-chip integration time) and Coadd (number of frames to add together before the file is saved). The Itime is set so that the background + object flux will be lower than the saturation level. The user can change filters and aperture wheel position by clicking on the icons in the lower left part of the window. The user can enter information about the object and observers, and specify the file name for the output images.  

\bibliography{main_ref}{}
\bibliographystyle{aasjournal}

\end{document}